\documentclass[iop,revtex4]{emulateapj}

\shorttitle{{\it Suzaku} observations of H 1743$-$322 in 2012 outburst}
\shortauthors{Shidatsu et al.}

\begin{document}

\title{Spectral and timing properties of the black hole X-ray binary H 1743$-$322\\ 
in the low/hard state studied with {\it Suzaku}}

\author{M. Shidatsu\altaffilmark{1}, Y. Ueda\altaffilmark{1}, 
S. Yamada\altaffilmark{2}, C. Done\altaffilmark{3}, T. Hori\altaffilmark{1}, 
K. Yamaoka\altaffilmark{4}, A. Kubota\altaffilmark{5}, T. Nagayama\altaffilmark{6}, 
Y. Moritani\altaffilmark{7}}
\email{shidatsu@kusastro.kyoto-u.ac.jp}

\altaffiltext{1}{Department of Astronomy, Kyoto University, Kitashirakawa-Oiwake-cho, 
Sakyo-ku, Kyoto 606-8502, Japan}
\altaffiltext{2}{Department of Physics, Tokyo Metropolitan University, Minami-Osawa 1-1, 
Hachioji, Tokyo, 192-0397, Japan}
\altaffiltext{3}{Department of Physics, University of Durham, South Road, Durham, DH1 3LE, UK}
\altaffiltext{4}{Solar-Terrestrial Environment Laboratory, Nagoya University, Furo-cho, 
Chikusa-ku, Nagoya, Aichi 464-8601, Japan}
\altaffiltext{5}{Department of Electronic Information Systems, Shibaura Institute of Technology, 
307 Fukasaku, Minuma-ku, Saitama 337-8570, Japan}
\altaffiltext{6}{Department fo Physics, Faculty of Science, Kagoshima University, 
1-21-35 Korimoto, Kagoshima, Kagoshima 890-0065}
\altaffiltext{7}{Department of Physical Sciences, Hiroshima University, Higashi-Hiroshima, 
Hiroshima 739-8526, Japan}

\begin{abstract}
We report on the results from 
{\it Suzaku} observations of the Galactic black hole
X-ray binary H 1743$-$322 in the low/hard state 
during its outburst in 2012 October. We appropriately take into account 
the effects of dust-scattering to accurately analyze the X-ray spectra.
The time-averaged spectra in the 1--200 keV band are dominated by a 
hard power-law component of a photon index of $\approx$ 1.6 with a
high-energy cutoff at $\approx$60 keV, which is well described with
Comptonization of the disk emission by hot corona.
We estimate the inner disk radius from the multi-color disk 
component, and find that it is 1.3--2.3 times larger than the radius 
in the high/soft state. This suggests that the standard disk was not 
extended to the innermost stable circular orbit. A reflection component 
from the disk is detected with $R = \Omega/2\pi \approx 0.6$ ($\Omega$ is 
the solid angle). We also successfully estimate the stable disk component 
in a way independent of the time-averaged spectral modeling, by analyzing
short-term spectral variability on the $\sim$1-sec timescale. A weak
low-frequency quasi-periodic oscillation (LF QPO) at 0.1--0.2 Hz is
detected, whose frequency is found to correlate with the X-ray
luminosity and photon index. This result may be explained by the 
evolution of the disk truncation radius.

\end{abstract}

\keywords{accretion, accretion disks --- black hole physics --- 
X-rays: binaries --- X-rays: individual(H~1743$-$322)}

\section{Introduction}
\setcounter{footnote}{0}
A large fraction of known Galactic black hole X-ray binaries 
(BHXBs) are highly transient sources. They are thought 
to be low-mass X-ray binaries, in which a stellar mass black 
hole accretes gas from a low-mass donor star via Roche-lobe overflow. 
They normally stay in a very faint state, but occasionally undergo 
dramatic outbursts, changing their X-ray luminosities in several orders 
of magnitude \citep[e.g.,][]{tan96} on a human timescale (typically 
$\approx$10 days to several months for the whole period of an outburst). 
Thus, they are excellent targets to study the physics 
of black hole accretion over a wide range of mass accretion rates.
Along with the increase and decrease of the X-ray luminosity, 
they exhibit several distinct ''states'' with different spectral 
and timing properties. The most canonical ones among them are 
so-called the ''high/soft state'' and the ''low/hard state'', which 
appears in relatively high and low luminosities 
($\sim 0.1 L_{\rm Edd}$\footnote{$L_{\rm Edd}$ is the Eddington luminosity: 
$4 \pi G m_{\rm p} c M_{\rm BH}/ \sigma_{\rm T}$, where $G$, $m_{\rm p}$, $c$, 
$M_{\rm BH}$, and $\sigma_{\rm T}$ represent the gravitational constant, 
proton mass, speed of light, black hole mass, and Thomson scattering 
cross-section, respectively.}, and $\lesssim 10^{-2} L_{\rm Edd}$, 
respectively; see e.g., \citealt{mcc06, don07} for a review).

In the high/soft state, the X-ray flux is dominated by the thermal 
emission from the optically-thick and geometrically-thin accretion 
disk \citep[the standard disk;][]{sha73}, and the observed spectrum 
is well reproduced with a multi-color disk model \citep{mit84}. 
The inner disk radius stays constant during the high/soft state, 
regardless of the significant change in X-ray luminosity 
\citep[e.g.,][]{ebi93, kub04, ste10, shi11a}. This indicates that 
the standard disk stably extends down to the 
innermost stable circular orbit (ISCO). In the low/hard state, 
BHXBs exhibit a hard, power-law shaped X-ray spectrum with a photon 
index of $< 2$ and an exponential cutoff at $\approx$100 keV. 
This spectral shape is generally described as unsaturated Compton 
scattering of the soft X-ray photons from the accretion disk 
by thermal electrons in hot inner flow or corona. Previous studies 
suggested that the standard disk is truncated in the low/hard state 
and eventually extends inward according to the increase of X-ray 
luminosity \citep{tom09,shi11b,yam13,kol14}. However, in what 
luminosity the standard disk reaches the ISCO is still controversial. 
Further study is necessary to fully understand the inner disk structure 
and its evolution in response to the change in mass accretion rate 
in the low/hard state.

Fast time variability of BHXBs gives clues to reveal the structure 
of the inner accretion disk and corona. In the low/hard state, the 
power density spectra (PDSs) are roughly described by a strong 
band-limited noise component, a flat-topped spectrum in the 
$\nu P_\nu$ form between the low- and high-frequency breaks. 
The low-frequency break moves to higher frequencies according 
to the rapid rise of the X-ray flux. This suggests that 
the break is associated with the truncation radius of the standard 
disk \citep{ing12}. Extracting the energy spectra of the noise 
component, \citet{axe05} found that the fast variability is produced 
by Comptonized emission from the vicinity of the black hole. 
Quasi-periodic oscillations (QPOs) are sometimes detected on 
the band-limited noise. Previous studies reported that the observed 
QPO frequencies are correlated to the spectral index and disk flux 
\citep[e.g.,][]{sob00, tit04, sha07}, 
implying that the QPOs, as well as the band-limited noise, are 
coupled with the inner disk structure.

H 1743$-$322 is a transient BHXB discovered by {\it Ariel-V} in 1977 
\citep{kal77}. Since then, it has displayed more than 10 outbursts, 
which were extensively observed by X-ray satellites. 
This source exhibits outbursts recurrently with an interval 
of $\approx$200 days \citep{shi12}. Some of them were ''failed'' 
outbursts, in which the high/soft state is missing 
\citep[e.g.,][]{cap09, che10}. 
High- and low-frequency QPOs were detected above 100 Hz and 
below 10 Hz in many observations \citep[e.g.,][]{hom03, tom03, hom05}. 
The X-ray spectra of H 1743$-$322 were also obtained many times in 
various states, including wide-band data up to $\sim$100 keV taken with 
{\it INTEGRAL} and {\it Suzaku} \citep{cap05, blu10}.
Ionized absorption lines, likely originated from disk winds, were 
discovered in the high/soft state and the steep power-law 
(or ''very high'') state in {\it Chandra} high-resolution spectra 
\citep{mil06}. This suggests that the source has a high 
inclination angle.

H 1743$-$322 was also observed in other wavelengths. 
Optical and near-infrared counterparts were identified by 
\citet{ste03} and \citet{bab03}, respectively. However, the 
high Galactic extinction has hampered deeper monitoring 
observations to measure the black hole mass dynamically. 
Bright radio flares likely associated with relativistic jets 
were detected in the early phase of the outbursts 
\citep[e.g.,][]{rup03}. In the 2003 outburst, bipolar large-scale 
X-ray jets were resolved with {\it Chandra} \citep{cor05}. Combining 
the {\it Chandra} image with the radio data obtained with the {\it VLA} 
during the same outburst, \citet{ste12} estimated the distance, the 
inclination angle, and the spin parameter $a$ of the black hole 
($a = cJ/GM_{\rm BH}^2$, where $J$ represents the angular momentum) 
as $8.5 \pm 0.8$ kpc, 
$75^\circ \pm 3^\circ$, and $0.2 \pm 0.3$, respectively. 
They were derived by modeling the trajectories of 
the bipolar jets with a symmetric kinematic model.

On 2012 September 24, the brightening of H 1743$-$322 was reported by 
\citet{shi12} with {\it Monitor of All-sky X-ray Image} ({\it MAXI}; 
\citealt{mat09})/Gas Slit Camera (GSC; \citealt{mih11}). About 2 weeks 
later, we triggered three sequential Target-of-Opportunity (ToO) 
observations with {\it Suzaku} \citep{mit07}, when the X-ray flux 
reached to the peak and started decreasing. Quasi-simultaneous 
observations in the near-infrared ($J$, $H$, and $K_{\rm S}$) and 
optical ($R$ and $i$) bands were also carried out with the 
{\it Infrared Survey Facility} ({\it IRSF}) 1.4m telescope and the 
{\it Kanata} 1.5m telescope at Higashi-Hiroshima Observatory, 
respectively. The source was not detected in any bands, however.

In this paper, we report the results of spectral and timing studies
using the {\it Suzaku} datasets. Section 2 describes the detail of the
{\it Suzaku} observations and the data reduction. In Section 3, we 
present the {\it Suzaku} light curves and power spectra. The spectral 
analysis and the results are described in Section 4. The infrared and 
optical observations are summarized in Section 5. Section 6 gives the 
discussion of the {\it Suzaku} results and Section 7 is the summary and
conclusions. Errors represent the 90\% confidence range for a single
parameter of interest, unless otherwise specified.
Throughout the article, we refer to the table by \citet{and89} as 
the Solar abundances. We assume the distance and inclination of 
H 1743$-$322 as those determined by \citet{ste12}, $D=8.5$ kpc 
and $i=75^\circ$.

\section{{\it Suzaku} Observations and Data Reduction}

\begin{deluxetable*}{lccc}
\tablecaption{Log of the {\it Suzaku} observations\label{tab_obslog_Suzaku}}
\tablewidth{0pt}
\tablehead{
\colhead{} & \colhead{Epoch-1} & \colhead{Epoch-2} & \colhead{Epoch-3} 
}
\startdata
ObsID & 407005010 & 407005020 & 407005030 \\
start time (UT) & 2012 Oct. 4 18:46:29 & 2012 Oct. 10 15:30:34 & 2012 Oct. 12 09:43:11 \\
end time (UT) & 2012 Oct. 5 14:35:16 & 2012 Oct. 11 14:04:14 & 2012 Oct. 13. 08:59:13 \\
XIS window & 1/4 & 1/4 & 1/4 \\
burst option & 1.0 sec & 1.0 sec & 1.0 sec\\
net exposure & & & \\
XIS & 21 ksec & 21 ksec & 21 ksec \\
HXD & 42 ksec & 42 ksec & 43 ksec 
\enddata
\end{deluxetable*}

\begin{figure}
\plotone{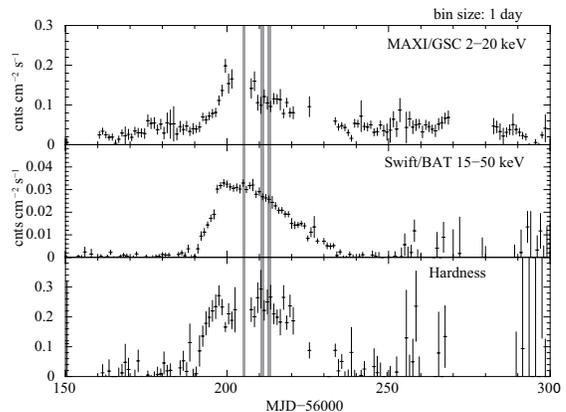}
\caption{Long-term light curves of H~1743$-$322 in 
the 2--20 keV band obtained with 
{\it MAXI}/GSC (top) and in the 15--50 keV band 
from {\it Swift}/BAT (middle),
and the hardness ratio calculated from these two light curves (bottom). The 
shadowed region indicate the {\it Suzaku} observations. MJD 56201 corresponds 
to 2012 October 1.\label{fig_longtermLC}}
\end{figure}

{\it Suzaku} ToO observations of H 1743$-$322 were performed on 2012 
October 4, 10, and 12, with a net exposure of $\approx$40 ksec in
each epoch. Figure~\ref{fig_longtermLC} plots the long-term X-ray 
light curves of {\it MAXI}/GSC and {\it Swift}/Burst Alert Telescope 
\citep[BAT;][]{bar05} and their hardness ratio, in which the {\it Suzaku} 
epochs are marked with shadowed regions. The first observation 
(hereafter ``Epoch-1'') is close to the peak of the BAT hard X-ray flux 
in the 15--50 keV band and the second and third ones (``Epoch-2'' and
``Epoch-3'', respectively) are in the decaying phase, where the fluxes
are by $\approx$20\% lower than that of Epoch-1.

{\it Suzaku} carries three currently available X-ray
charge-coupled-devise (CCD) cameras named the X-ray Imaging Spectrometer
(XIS), which detect X-ray photons of 0.2--12 keV, and a hard X-ray
sensor called Hard X-ray Detector (HXD) composed of PIN diodes and GSO
scintillators covering the 10--70 keV and 40--600 keV bands,
respectively. The XIS consists of frontside-illuminated cameras
(FI-XISs; XIS-0 and XIS-3) and a backside-illuminated one (BI-XIS;
XIS-1), which is more sensitive to soft X-rays below $\approx$1.5 keV
than FI-XISs. In our observations, the 1/4 window mode and the 1.0 sec
burst option were employed for both FI- and BI-XISs to avoid pile-up
effects. A summary of the {\it Suzaku} observations is given in
Table~\ref{tab_obslog_Suzaku}.

We utilized HEAsoft 6.13 and {\it Suzaku} Calibration Database (CALDB)
released on 2013 March 5 for the reduction and analysis of our data. We
started with the ``cleaned'' events (whose process version is 2.8.16.34)
provided by the {\it Suzaku} team and followed the standard procedure
for the data reduction.
Source photons were extracted from a circular region with a radius of
110'' centering at the target position. Background events were collected
within a circle with a radius of 120'' taken in a blank-sky area outside
of the region for source extraction. The XIS response matrix and
ancillary response files were generated by {\tt xisrmfgen} and {\tt
xissimarfgen} \citep{ish07}, respectively.  The non X-ray background
(NXB) was subtracted from the HXD data by using the modeled NXB files
produced by the HXD
team\footnote{http://www.astro.isas.ac.jp/suzaku/analysis/hxd/pinnxb/tuned/
for PIN, and http://www.astro.isas.ac.jp/suzaku/analysis/hxd/gsonxb/ for
GSO.}.  In the spectral analysis, {\tt
ae\_hxd\_pinxinome11\_20110601.rsp} was used as the response file for
PIN, and {\tt ae\_hxd\_gsoxinom\_20100524.rsp} and {\tt
ae\_hxd\_gsoxinom\_crab\_20100526.arf} were as those for GSO.

Because H 1743$-$322 is located in a crowded region close to the center of 
our Galaxy ($l,b = 357^\circ.26,-1^\circ.83$), the HXD spectra could be 
contaminated with the Galactic diffuse X-ray background (GXB) 
and nearby bright point sources. However, we find the 
GXB contributes only $\approx$1.5\% of the total flux in the PIN bandpass, 
and far smaller in that of GSO, which are negligible.
To estimate these values, we assumed the GXB spectrum at 
$(l, b) = (28^\circ.5, 0^\circ.2)$ shown in \citet{kub10} and converted 
the fluxes in the PIN and GSO energy ranges to those at the position 
of H 1743$-$322 using the spatial profile of the GXB in the 16--70 keV band
obtained with {\it INTEGRAL}/IBIS \citep{kri07}. The Cosmic X-ray 
background (CXB) is not subtracted from the HXD data, because its 
contribution is even smaller than that of the GXB. We examined the 
contamination of nearby sources using the {\it Swift}/BAT 70 month catalog 
\citep{bau13}, considering the detector's transmission functions \citep{tak07}. 
We also checked the variability of the sources through the {\it MAXI}/GSC and 
{\it Swift}/BAT public light curves. The nearby sources were found to 
contribute only $< 1\%$ to the observed PIN flux and $\approx$1\% 
to that of GSO, both of which are ignorable. 

\begin{figure*}
\plotone{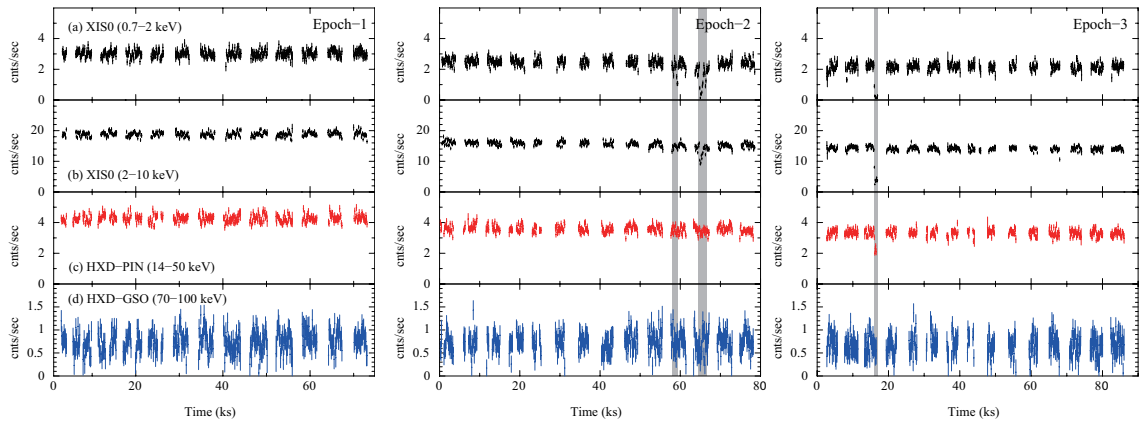}
\caption{{\it Suzaku} XIS and HXD light curves of Epoch-1 (left), Epoch-2 
(middle), and Epoch-3 (right). The shadowed regions are dipping periods.
\label{fig_suzakuLC}}
\end{figure*}

\section{X-ray Light Curves and Timing Properties}

\begin{figure}
\epsscale{1.0}
\plotone{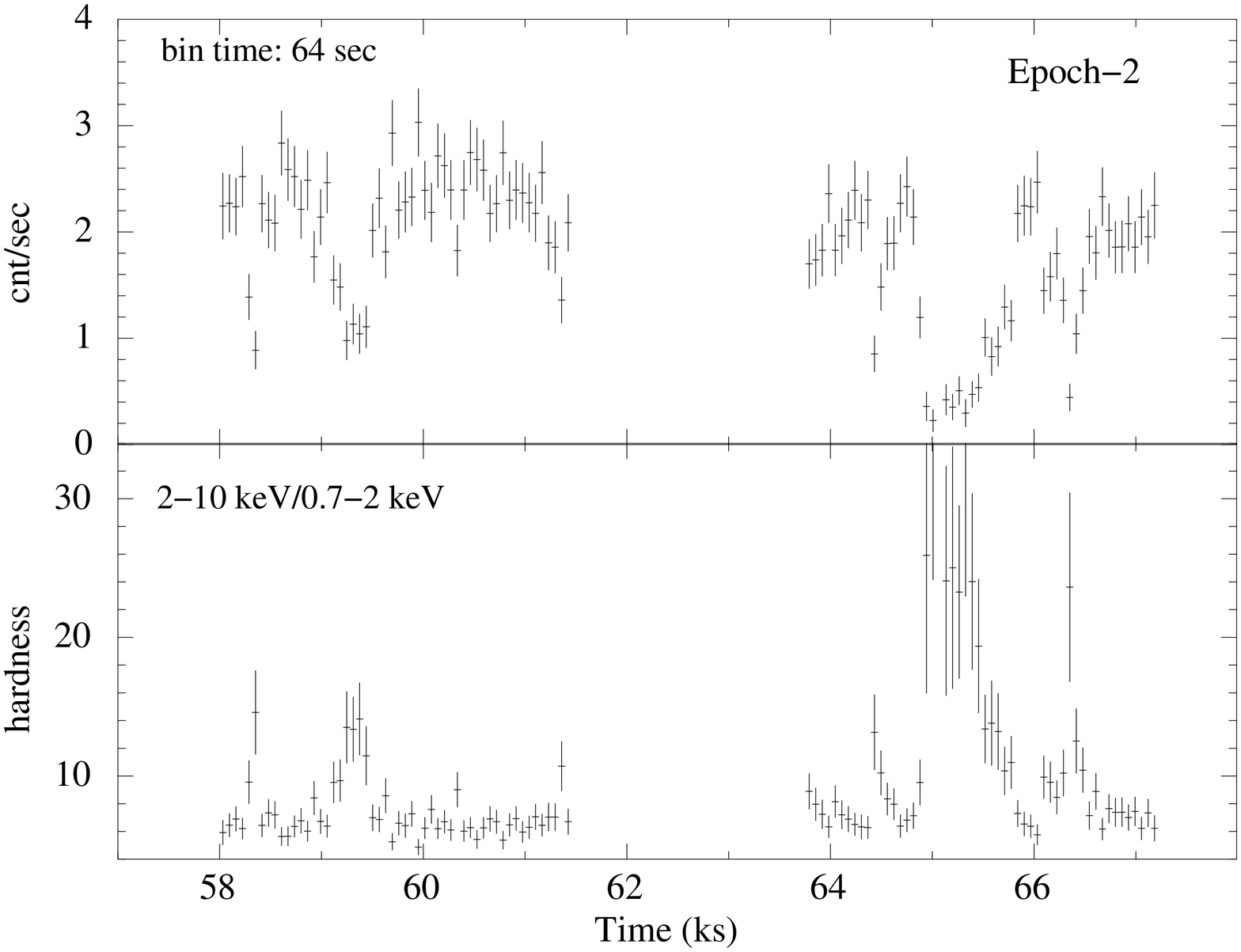}
\caption{{\it Suzaku} light curve and hardness ratio in Epoch-2 focused 
on the dipping periods. (Upper) the XIS-0 light curve in the 0.7--2 keV band 
binned in 64 sec. (Lower) The hardness ratio between the 2--10 keV and 0.7--2 
keV bands. Similar hardening can also be seen in the dipping period of Epoch-3.
\label{fig_lc_dip}}
\end{figure}

Figure~\ref{fig_suzakuLC} presents the {\it Suzaku} light curves in the
three epochs. They sometimes display significant drops of the soft X-ray
flux in Epoch-2 and Epoch-3. These are likely ``absorption dips'', which
are thought to be caused by the obscuration of the central X-ray source
by the outer edge of the accretion disk hit by the accretion stream from
the companion star. As noticed in Figure~\ref{fig_lc_dip}, the 
hardness ratio increases in these dipping periods. This indicates that 
the dips are caused by photo-electric absorption, although the too 
short dipping duration does not allow us to perform precise time-resolved 
spectral analysis.
The presence of similar absorption dips in H 1743$-$322 was also 
reported in previous observations \citep[e.g.,][]{hom05}. We define
the shadowed regions in Fig.~\ref{fig_suzakuLC} as the dipping periods,
in which the count rate in the 0.7--2 keV band is $\gtrsim$30\% smaller
than the averaged rate in the individual epochs, and exclude them in the
following analysis.

\begin{figure}
\epsscale{1.0}
\plotone{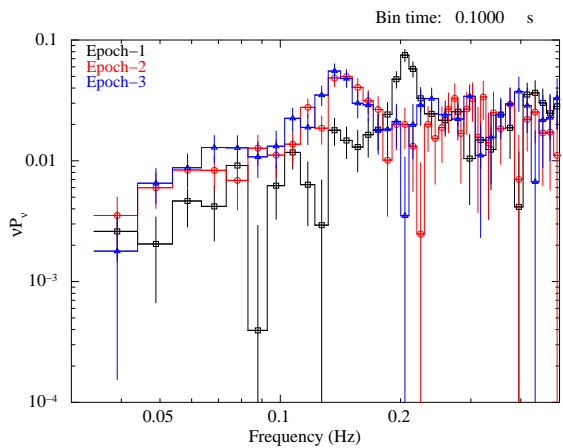}
\caption{Normalized power density spectra (PDSs) in the three epochs created from 
the PIN light curves with 0.1 sec bins. They are normalized in the way that their 
integral gives the squared root mean squared fractional variability. White noise is 
subtracted from all the PDSs.\label{fig_psd}}
\end{figure}

Figure~\ref{fig_psd} plots the normalized power density spectra
(PDSs) in the three epochs obtained with the PIN data.
The PDSs are dominated by a strong noise
components with a normalized power of $\approx 10^{-2}$ (rms/mean)$^2$
in the $\nu P_{\nu}$ form, consistent with those of typical BHXBs in the
low/hard state \citep[see e.g.,][]{mcc06}. A weak low frequency QPO is
also detected in each epoch. Fitting it with a Gaussian model, we
estimate the central frequency of the QPO as $0.206 \pm 0.003$ Hz in
Epoch-1, which become 1.4 times lower in Epoch-2 and Epoch-3 ($0.147 \pm
0.004$ Hz and $0.141 \pm 0.004$ Hz, respectively) as the {\it Swift/BAT}
flux is decreased by $\approx$ 20\%.
This QPO is detected in the XIS PDSs as well.

\section{Analysis of {\it Suzaku} Spectra}

\begin{figure}
\plotone{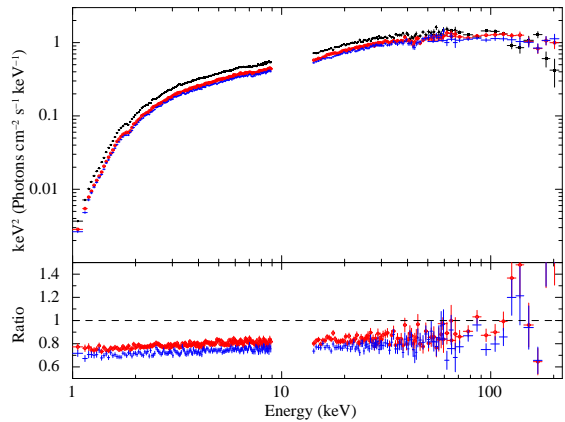}
\caption{The time-averaged spectra in Epoch-1 (black, filled square), Epoch-2 (red, open circle), 
and Epoch-3 (blue, cross). The spectral ratios of Epoch-2 and Epoch-3 with respect to Epoch-1 
are plotted in the lower panel.\label{fig_spec}}
\end{figure}

We create the time-averaged spectra of the three epochs and analyze them
separately on XSPEC version 12.8.0. Considering the signal-to-noise
ratio and the reliability of the calibration, we use the data within
1--9 keV, 1--8 keV, and 14--70 keV for FI-XISs, BI-XIS, and PIN,
respectively. For the GSO data, we only consider the 50--130 keV (Epoch-1),
50--220 keV (Epoch-2), and 50--200 keV (Epoch-3) bands, 
above which the signal 
levels are overcome by the systematic errors in the simulated
background \citep[$\lesssim$1\% of the total count rate;][]{fuk09}.  The
spectra and responses of FI-XISs (i.e., XIS-0 and XIS-3) are combined to
improve statistics. The XIS data of 1.7--1.9 keV are discarded to
avoid the uncertainties in the responses at energies around the
complex instrumental Si-K edge. To account for possible calibration
uncertainties, we add a 1\% systematic error to every spectral bin in the
XIS and HXD data, following \citet{mak08}. We find that the final 
best-fit parameters are the same within their 90\% confidence ranges 
even if we assign 0\%, 2\%, and 3\% systematic errors, and that the 
confidence ranges themselves are increased only by $\lessapprox 10$\% 
from the case of 0\% systematic error to that of 3\%. 
We confirm that the conclusions are not affected.
The cross-normalization of the HXD with respect to FI-XISs was fixed at
1.16\footnote{http://www.astro.isas.ac.jp/suzaku/doc/suzakumemo/suzaku\\memo-2008-06.pdf},
while that of BI-XIS is left free. We examine the pileup effects in the
XIS data using the software {\tt aepileupcheckup.py}, which estimates
the pileup fraction at various radii from the center of the XIS point
spread function \citep{yam12}. We find that the pileup fraction is
negligibly small in all observations, less than 2\% even at the cores of
the PSFs. Figure~\ref{fig_spec} shows the time-averaged 
spectra for the three epochs in the $\nu F_\nu$ form.

\subsection{Dust-scattering Effects}

In observations of heavily absorbed sources (with $N_{\rm H} \gtrsim
10^{22}$ cm$^{-2}$), it must be borne in mind that dust scattering 
in addition to photo-electric absorption can significantly affect the
observed soft X-ray spectra \citep{ued10,hor14}. Diffuse interstellar
(and circumbinary) dust can scatter-out the X-ray photons from the line
of sight and at the same time, scatter-in the photons emitted in
different directions. Consequently, we see a ''dust halo'' around a
point source. A dust halo was indeed resolved in GX 13$+$1
\citep{smi02}, which has almost the same column density ($N_{\rm H}
\approx 2 \times 10^{22}$ cm$^{-2}$) as our target. When we use 
data that do not fully cover the dust halo, 
those two scattering effects are not canceled out. In this
case, the observed spectra, particularly in the soft X-ray band, are
changed from what we would expect if there were no dust-scattering. This
must be taken into account for accurate spectral modeling. Even if all
the scattered photons are collected, the scattered-in and scattered-out
components no longer compensate each other in variable sources, because
of the difference in arrival time between the scattered photons and
those taking the straight path to the observer.  In our case, the
typical delay time of the scattered photons in the extraction region for
the XIS is estimated to be $\sim 1.3$ day, by assuming a distance of 8.5
kpc. We confirm for all the {\it Suzaku} epochs that the flux
variability of H 1743$-$322 on that timescale is low enough to ignore
the variability of the scattered-in component, as in the case of
\citet{hor14}.

Following \citet{ued10} and \citet{hor14}, we utilize {\tt Dscat}, a
local multiplicative model implemented in XSPEC, to account for the
effects of dust scattering in the spectral analysis. {\tt Dscat}
estimates both scattering-in and -out components from the scattering
fraction (the fraction of the scattered-in photons included in the data)
and the hydrogen column density, by assuming the dust scattering cross
section for $R_{\rm V} = 3.1$ calculated by \citet{dra03}. We fix the
scattering fraction at 1.0 (i.e., all the scattered photons are
included) for the HXD, which has sufficiently large field of view to
cover the whole dust halo. For the XIS, we adopt the same value (0.65)
as \citet{hor14}, because we used the same region as theirs to extract
the source photons from the XIS data. In the following analysis, the
column density of the {\tt Dscat} model is always linked to that of the
interstellar photo-electric absorption, for which we employ {\tt phabs}
\citep{bal92}. We find that $\approx$30\%, $\approx$8\%, and
$\approx$3\% of the intrinsic flux from H1743$-$322 are reduced at 1
keV, 3 keV, and 5 keV by the effects of dust scattering,
respectively\footnote{Here $N_{\rm H} = 2 \times 10^{22}$ cm$^{-2}$ is
assumed.}.

\subsection{Modeling Time-averaged Spectra}

As noticed in Fig.~\ref{fig_spec}, the overall spectral shapes in the
three epochs are very similar to one another, although the Epoch-2 and
Epoch-3 spectra are slightly harder than that of Epoch-1. All of them
are roughly characterized by a hard power-law component with a photon
index of $\approx 1.6$, suggesting that the source is in the low/hard
state during the {\it Suzaku} observations. The hydrogen column density
is estimated to be $\approx$2.0 $\times 10^{22}$ cm$^{-2}$ for the
photo-electric absorption in cold interstellar medium. This value is
well within those obtained in previous studies of H 1743$-$322
\citep[1.6--2.3 $\times 10^{22}$ cm$^{-2}$; e.g.,][]{cap09,mil06}. A 
high-energy rollover can be found in 50--100 keV, which is likely to 
correspond to the electron temperature of the thermal corona. No 
significant emission and absorption lines can be seen in the spectra.

\begin{figure}
\plotone{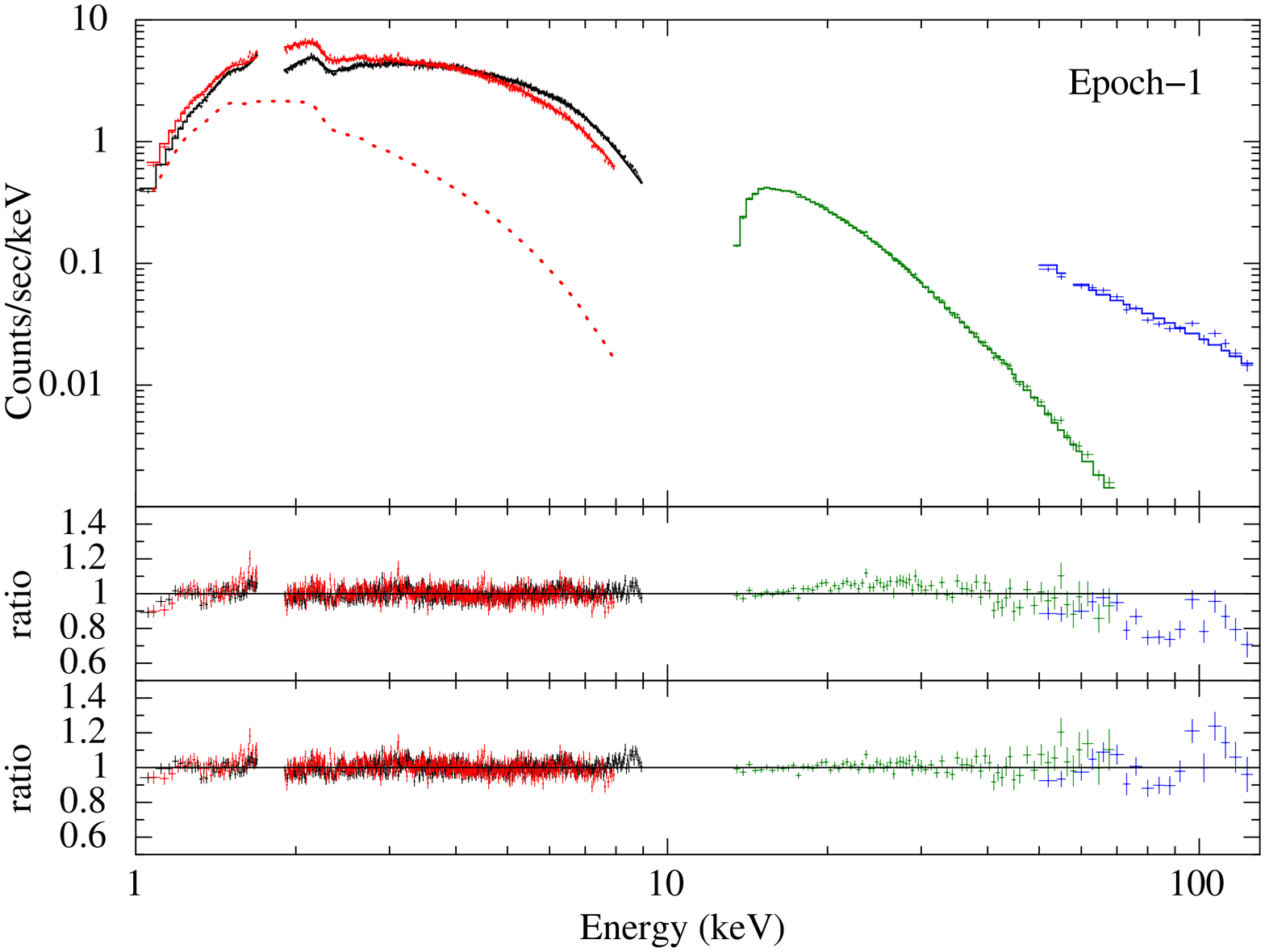}
\plotone{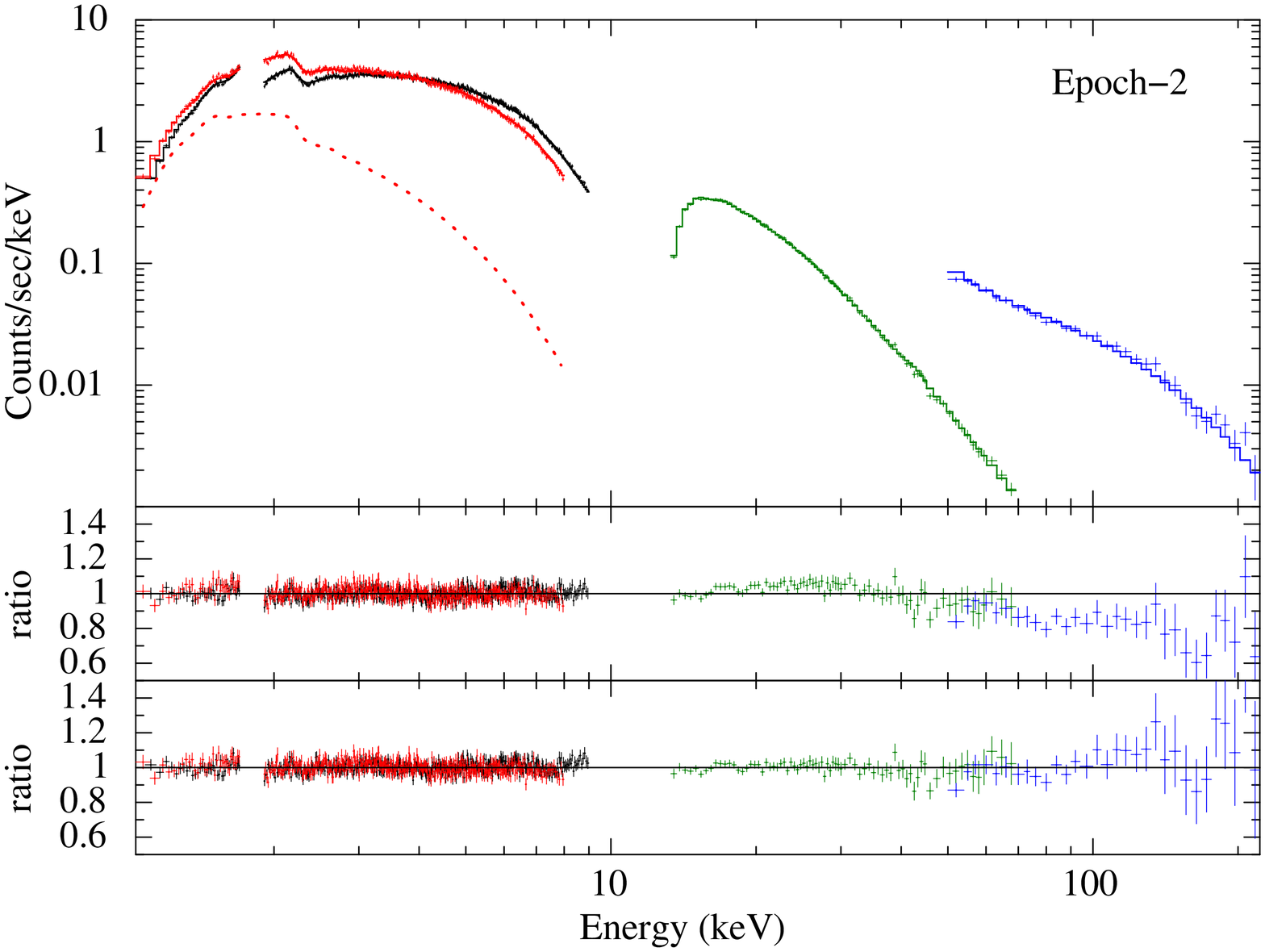}
\plotone{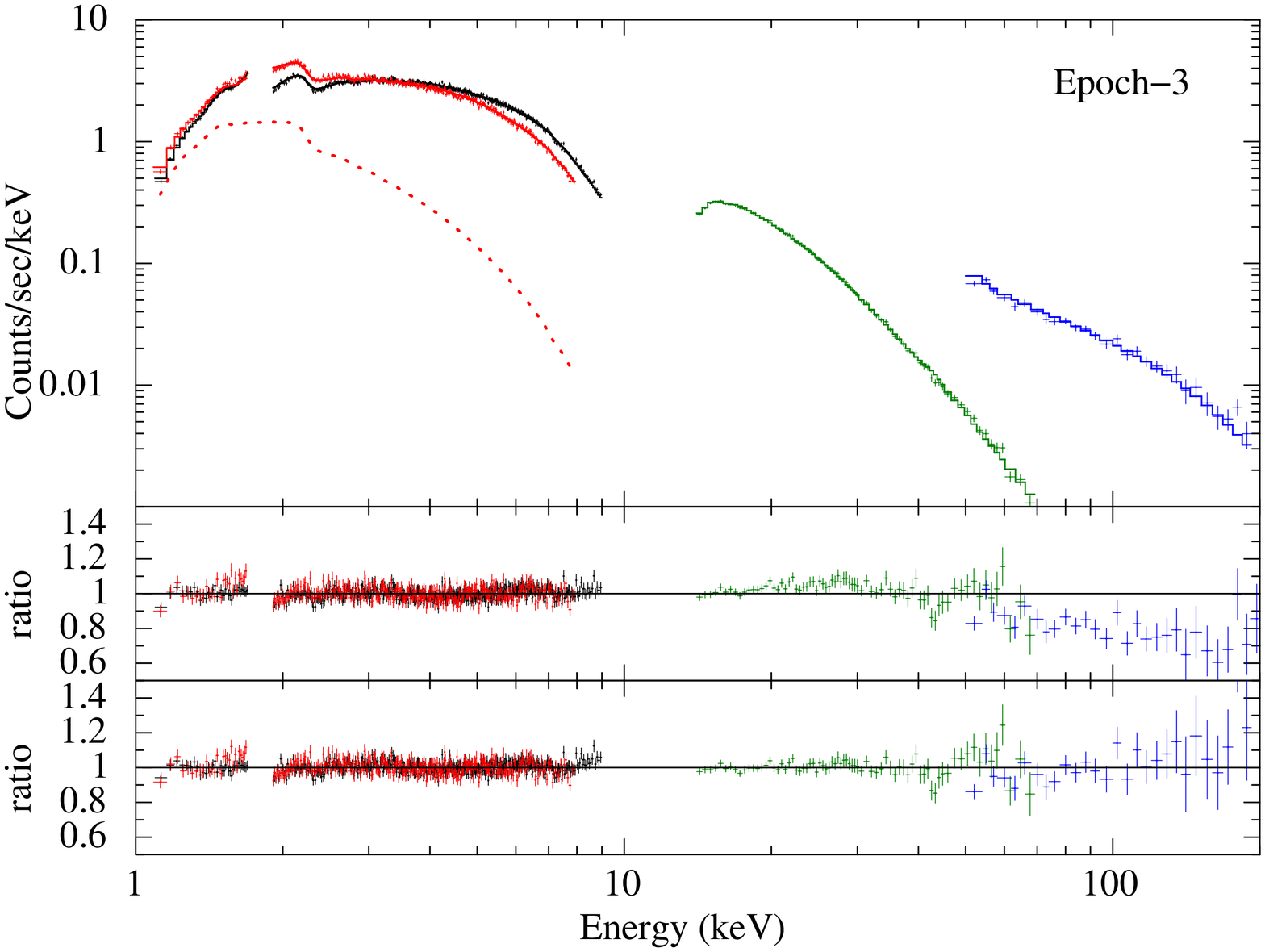}
\caption{The time-averaged spectra of Epoch-1 (top), Epoch-2 (middle), 
and Epoch-3 (bottom) fitted with the {\tt nthcomp} model. The black 
(below 9 keV), red (below 8 keV), green (14--70 keV), and blue 
(above 50 keV) points correspond to the XIS-0$+$XIS-3, XIS-1, PIN, and 
GSO data, respectively. The contribution of the scattered flux to the 
XIS-1 data is shown in red dotted line. The data/model ratio for a single 
{\tt nthcomp} model and the final best-fit model are plotted in the second 
and third panels, respectively.\label{fig_specfit}}
\end{figure}

We first fit the spectra with the {\tt nthcomp} model \citep{zdz96,
zyc99} with an interstellar absorption. The {\tt nthcomp} model
calculates a thermal Comptonization spectrum parametrized by a photon
index, an electron temperature of the Comptonizing corona ($kT_{\rm
e}$), and a seed-photon temperature. We could not constrain 
$kT_{\rm e}$
from a joint fit with the XIS and HXD spectra, due to much
poorer statistics of the GSO data compared with those of the XIS that
dominate the total $\chi^2$ statistics.
We therefore determine $kT_{\rm e}$ only from 
the GSO spectra\footnote{Here
the seed temperature is fixed at 0.1 keV to constrain $kT_{\rm e}$ so
that it does not affect the spectral shape in the GSO band.}, and fix 
it at the best-fit value in fitting together with the XIS and PIN data. 
From the joint fit of the
XIS$+$HXD spectra, we obtain $\chi^2/{\rm d.o.f.} = 1690/1158$
(Epoch-1), $1253/994$ (Epoch-2), and $1237/904$ (Epoch-3), which are far
from acceptable. As can be seen in the middle panel in
Fig.~\ref{fig_specfit}, significant residuals remain in the hard X-ray
band above 10 keV. A convex-shaped structure can be found, which is
likely originated from the reflection of Comptonized photons on the
disk.

We next apply a more sophisticated model to the spectra, considering the
general picture of the low/hard state. We use {\tt diskbb} model
\citep{mit84} as the direct emission from the standard disk and the {\tt
nthcomp} model as the Comptonization. The seed photon of the {\tt
nthcomp} component is assumed to be the multi-color disk emission, and
the seed temperature is tied to the inner disk temperature of the {\tt
diskbb} component. To account for the reflection component, the {\tt
ireflect} model \citep{mag95} is convolved to {\tt nthcomp}. This model
calculates a reflection spectrum using the ionization parameter ($\xi =
L_{\rm X}/(nR^2)$, where $L_{\rm X}$, $n$, and $R$ stand for the
incident X-ray luminosity, the electron number density and the distance
of reflector from the X-ray source, respectively), the temperature
($T_{\rm disk}$), inclination angle ($i$), and solid angle ($\Omega/2
\pi$) of the reflector. We fix $T_{\rm disk}$ at $30000$ K, which cannot
be constrained from the data.
The {\tt ireflect} model does not include any emission lines accompanied
by the reflection continuum. We thus combine a Gaussian component to
account for the iron K-$\alpha$ emission line, which is normally seen in
the low/hard state spectra most significantly. Considering numerical
studies \citep[e.g.,][]{mat91}, we link $\Omega/2 \pi$ of {\tt ireflect}
and the normalization of the Gaussian to keep the equivalent width with
respect to the reflection continuum at 1.0 keV. 
We note that the equivalent width becomes more than an order of 
magnitude smaller than what is expected in the numerical calculations, 
when we unlinked the normalization of the Gaussian and the solid angle of 
{\tt ireflect}.
The convolution model {\tt kdblur} \citep{lao91} is also incorporated to
model the relativistic blurring by the accretion disk around the 
black hole. This model uses the index $\alpha$ describing the radial
dependence of the emissivity $\epsilon$ ($\propto r^{-\alpha}$), inner
and outer radii of the accretion disk, and the inclination angle. Here
we assume $\alpha = 3$ (i.e., a flat disk) and fix the inner radius at
$10 R_{\rm g}$ (where $R_{\rm g}$ is the gravitational radius, $GM_{\rm
BH}/c^2$), and the outer radius at $R_{\rm out} = 400$, which is the
maximum value of the {\tt kdblur} model. The integrated fitting model 
is described as 
{\tt phabs*Dscat*(diskbb+kdblur*(ireflect*nthcomp[re\\f]+gauss)+nthcomp[C])}, 
where {\tt nthcomp[ref]} and {\tt nthcomp[C]} represent the parts of 
the Comptonized component that is and is not reflected on the disk, 
respectively. 
Because {\tt ireflect} and {\tt kdblur} are convolution models, we 
extend the energy range to 0.01--1000 keV.

\begin{figure}
\plotone{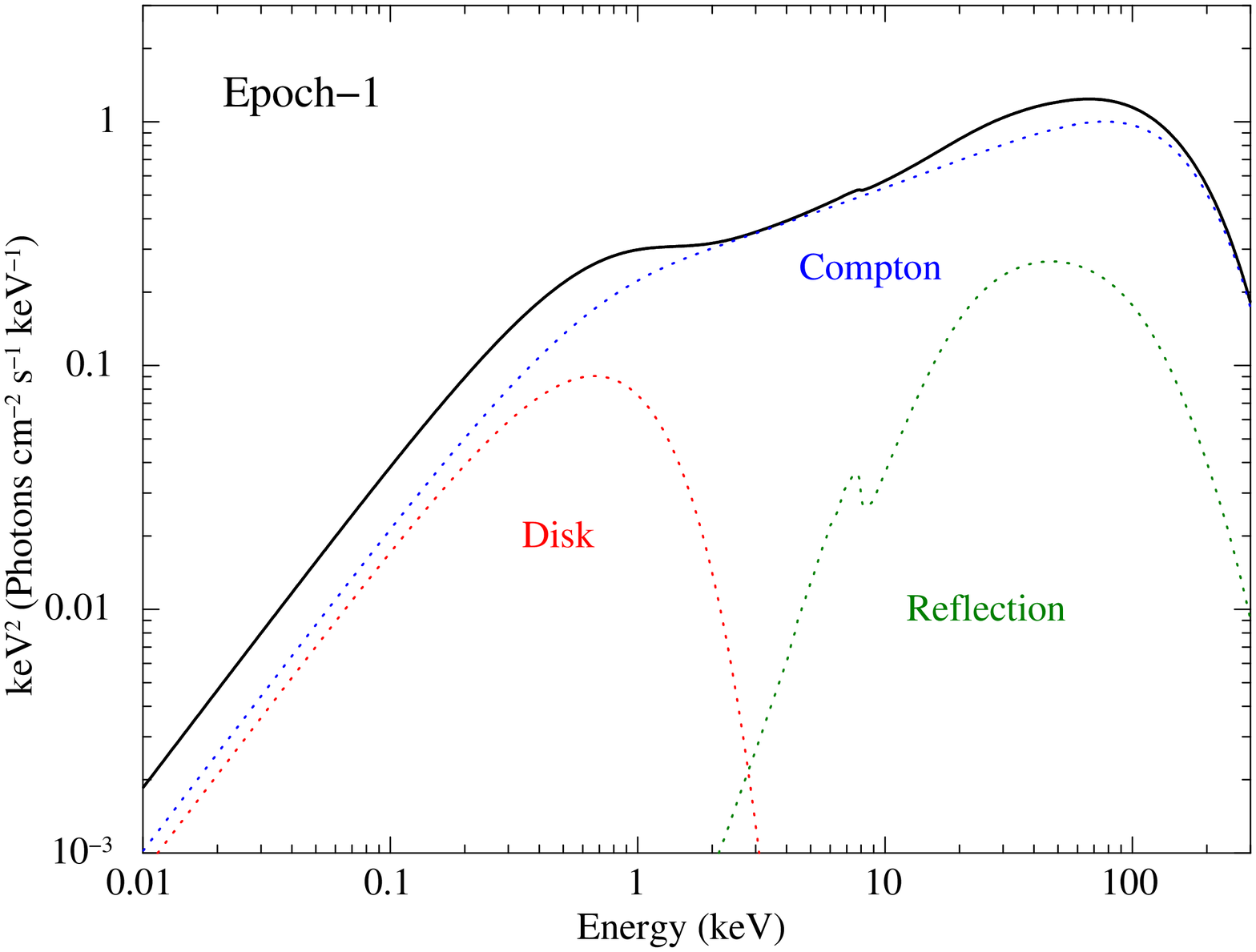}
\plotone{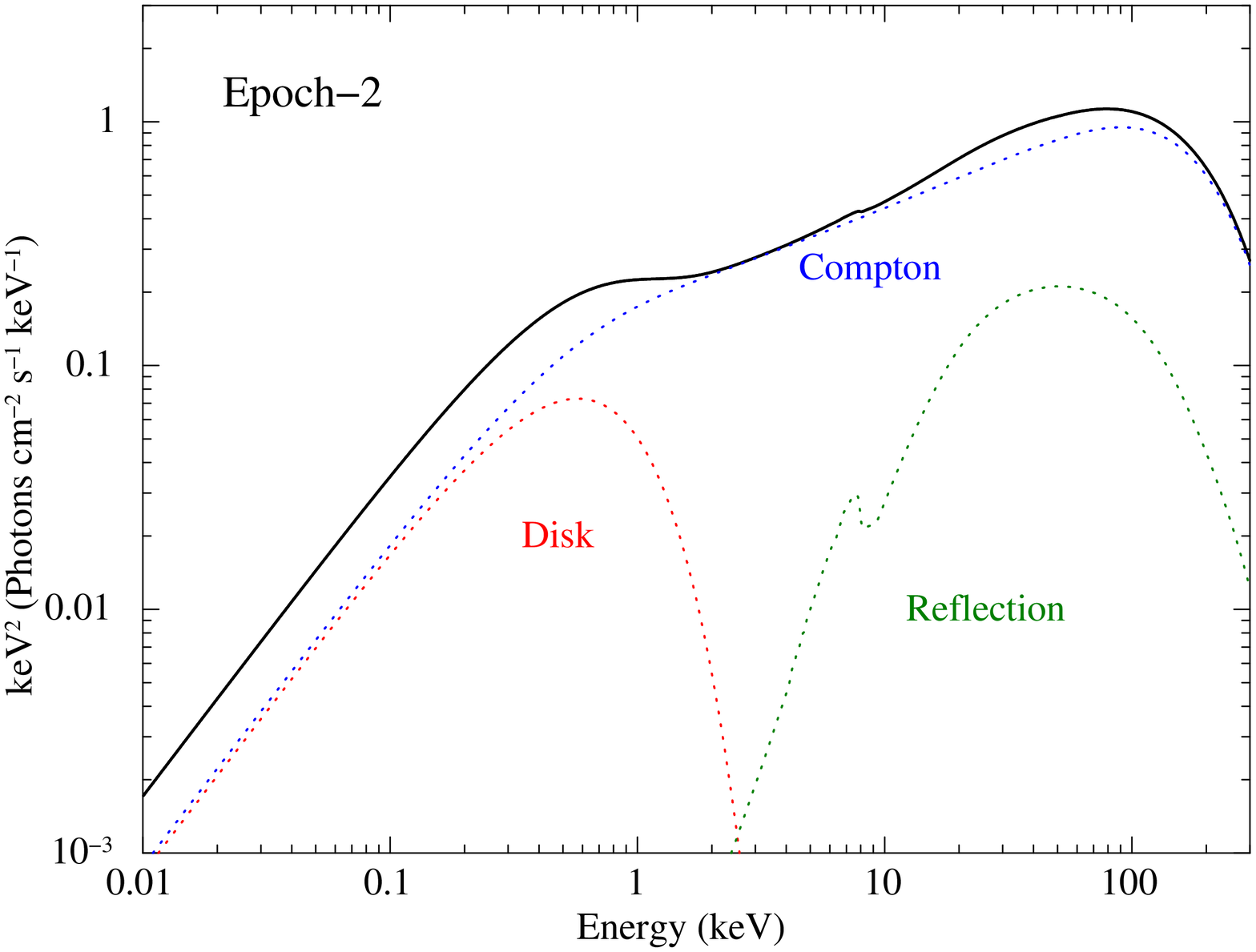}
\plotone{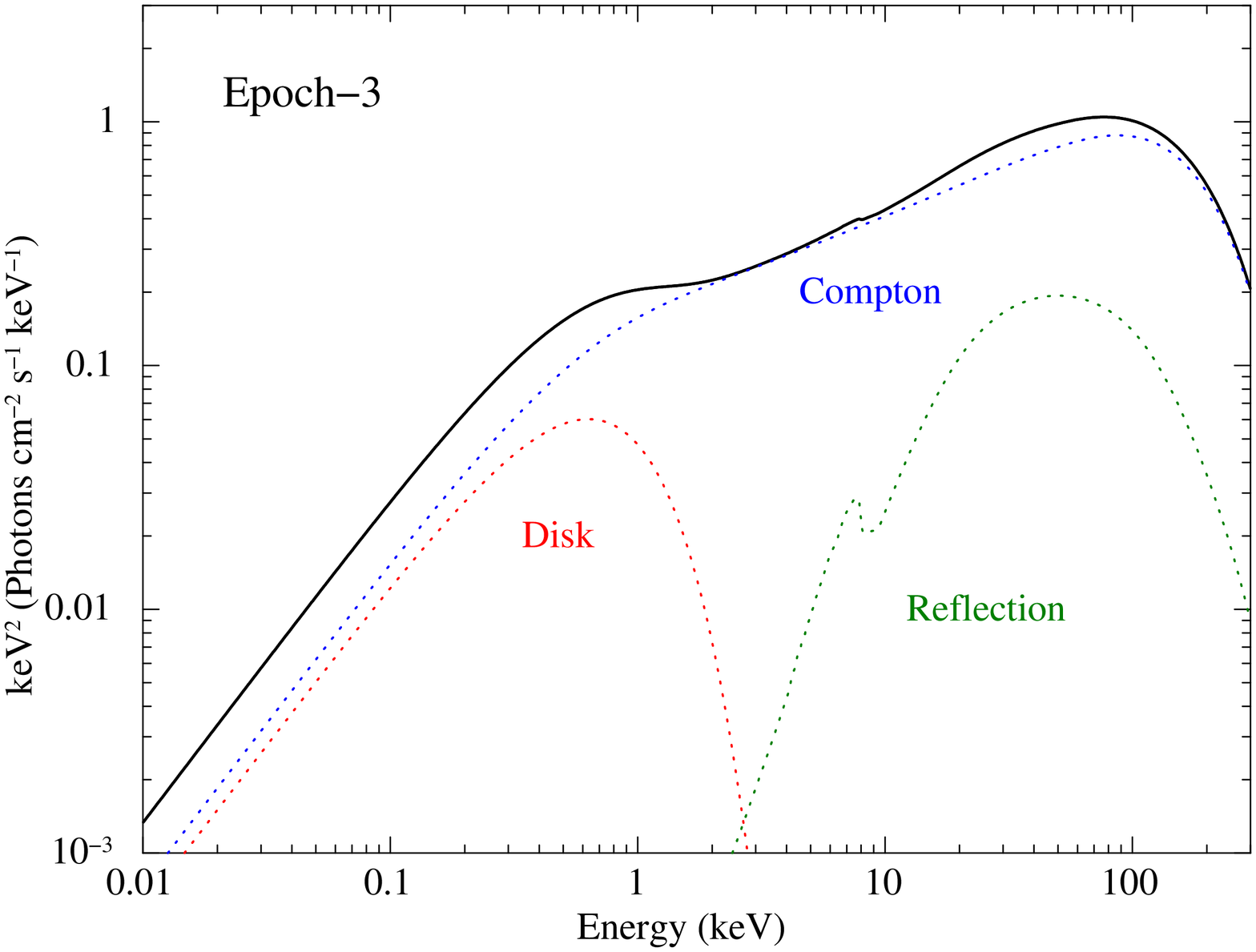}
\caption{The best-fit models in Epoch-1 (top), Epoch-2 (middle), and Epoch-3 
(bottom) plotted in the $\nu F_\nu$ form.
\label{fig_specmodel}}
\end{figure}

The quality of fit is remarkably improved compared with that of the
single {\tt nthcomp} model, yielding $\chi^2/{\rm d.o.f.} = 1347/1155$
(Epoch-1), $984/991$ (Epoch-2), and $995/901$ (Epoch-3). 
The spectra and the best-fit models in the individual epochs are plotted
in Figure~\ref{fig_specfit} and Figure~\ref{fig_specmodel},
respectively. The resultant parameters are given in
Table~\ref{tab_specfit}.
The inclusion of the relativistic blurring model to the reflection 
component makes a significant statistical improvement ($\Delta \chi^2 
=$ 70--90, where the degrees of freedom are not changed). When the 
{\tt kdblur} component is excluded, we obtain $\chi^2/{\rm d.o.f.} = 
1435/1155$, $1061/991$, and $1075/901$, for Epoch-1, -2, and -3, 
respectively. Conversely, the addition of a narrow Gaussian 
component at 6.4 keV with a 1-$\sigma$ width of 10 eV does not improve the 
quality of fit.

Assuming an isotropic Comptonizated corona and the conservation of the total 
number of photons from the accretion disk, the inner disk radius can be 
derived from the formula in \citet{kub04};
\begin{eqnarray}
F^p_{\rm disk}+F^p_{\rm thc}2 \cos i &= 0.0165 \left[ \frac{r_{\rm in}^2\cos i}{(D/10\mbox{ kpc})^2}\right] 
\left( \frac{T_{\rm in}}{1\mbox{ keV}} \right)^3 \nonumber \\
& \mbox{ photons } {\rm s}^{-1} \mbox{ }{\rm cm}^{-2}, \label{eq_Rin}
\end{eqnarray}
where $F^p_{\rm disk}$ and $F^p_{\rm thc}$ are the 0.01--100 keV photon 
fluxes of the disk and thermal Comptonized components, respectively. 
Extending the best-fit model to that energy range, we estimate 
$F^p_{\rm disk} = 0.699$, $0.646$, and $0.489$ photons cm$^{-2}$ sec$^{-1}$ 
and $F^p_{\rm thc} = 1.27$, $1.06$, and $0.918$ photons cm$^{-2}$ 
sec$^{-1}$ in order from Epoch-1 to Epoch-3. The inner disk radii are 
thus $r_{\rm in}=101^{+10}_{-9}$ $D_{8.5} (\cos i/\cos 75^\circ)^{-1/2}$ km, 
$r_{\rm in}=119^{+24}_{-15}$ $D_{8.5} (\cos i/\cos 75^\circ)^{-1/2}$ km, and 
$r_{\rm in}=92^{+19}_{-10}$ $D_{8.5} (\cos i/\cos 75^\circ)^{-1/2}$ km (where 
$D_{8.5}$ represents the distance in units of 8.5 kpc). Here we only 
include the 90\% confidence range of the inner disk temperature to 
estimate the uncertainties of the radii. Multiplying a representative 
correction factor (1.19) of boundary condition and spectral hardening 
\citep{kub98}, the actual radii are calculated to be 
$R_{\rm in} = 120^{+12}_{-11}$ $D_{8.5} (\cos i/\cos 75^\circ)^{-1/2}$ km, 
 $R_{\rm in} = 141^{+28}_{-17}$ $D_{8.5} (\cos i/\cos 75^\circ)^{-1/2}$ km, and 
$R_{\rm in} = 110^{+23}_{-12}$ $D_{8.5} (\cos i/\cos 75^\circ)^{-1/2}$ km for 
Epoch-1, -2, and -3, respectively.

The inner radius in the high/soft state is $R_{\rm in} = 64 \pm 2$ 
$D_{8.5} (\cos i/\cos 75^\circ)^{-1/2}$ km \citep{che10}, which 
is estimated from the direct disk emission without Comptonization. 
This value is taken from the first two rows of Table 2 in \citet{che10} 
(which are originally from \citealt{mcc09} and \citealt{cap09}). 
The fraction of the power-law component in the 2--20 keV flux is 
less than 6\% for these two results and therefore $R_{\rm in}$ 
is changed by only a few \% by including the Comptonized disk 
photons. 
The inner radii obtained from the {\it Suzaku} data are 1.3--2.3 times 
larger than the value in the high/soft state, suggesting that the 
standard disk do not extend to the ISCO during the low/hard state 
observations.

\begin{deluxetable*}{lcccc}
\tablecaption{Best-fit parameters of the {\tt nthcomp} model \label{tab_specfit}}
\tablewidth{0pt}
\tablehead{
\colhead{Component} & \colhead{Parameter} & \colhead{Epoch-1} & \colhead{Epoch-2} 
& \colhead{Epoch-3} 
}
\startdata
phabs & $N_{\rm H}$ ($10^{22}$ cm$^{-2}$) & $2.01^{+0.04}_{-0.06}$ & $1.99 \pm 0.04$ 
 & $1.99^{+0.05}_{-0.06}$ \\
diskbb & $kT_{\rm in}$ (keV) & $0.28 \pm 0.02$ & $0.24^{+0.03}_{-0.02}$ & $0.27^{+0.02}_{-0.03}$ \\
& norm & $2.1^{+0.8}_{-1.0} \times 10^{3}$ & $3.1^{+2.7}_{-1.3} \times 10^{3}$ 
 & $1.7^{+1.3}_{-0.8} \times 10^{3}$ \\
nthcomp & $\Gamma$ & $1.668^{+0.009}_{-0.005}$ & $1.633^{+0.009}_{-0.008}$ & $1.629 \pm 0.009$ \\
 & $kT_{\rm e}$ (keV)\tablenotemark{a} & $< 35$ 
& $68^{+68}_{-20}$ & $61^{+144}_{-21}$ \\
 & norm & $0.223^{+0.004}_{-0.006}$ & $0.174 \pm 0.003$ & $0.157 \pm 0.004$ \\
ireflect & $\Omega/2 \pi$ & $0.65^{+0.08}_{-0.12}$ & $0.56 \pm 0.07$ & 
$0.55^{+0.08}_{-0.07}$ \\
 & $\xi$ (erg cm sec$^{-1}$) & $< 6$ & $ < 31$ & $ < 68$ \\
 & $T_{\rm disk}$ (K) & 30000 (fixed) & 30000 (fixed) & 30000 (fixed) \\
kdblur & $\alpha$\tablenotemark{b} & 3 (fixed) & 3 (fixed) & 3 (fixed) \\ 
 & $R_{\rm in}$ ($R_{\rm g}$) & 10 (fixed) & 10 (fixed) & 10 (fixed) \\
 & $R_{\rm out}$ ($R_{\rm out}$) & 400 (fixed) & 400 (fixed) & 400 (fixed) \\
 & $i$ (deg) & 75 (fixed) & 75 (fixed) & 75 (fixed) \\
gauss\tablenotemark{c} & $E_{\rm cen}$ (keV) & 6.4 (fixed) & 6.4 (fixed) & 6.4 (fixed) \\
 & $\sigma$ (keV) & 0.01 (fixed) & 0.01 (fixed) & 0.01 (fixed) \\
observed flux & 1--10 keV (ergs cm$^{-2}$ sec$^{-1}$) & $9.3 \times 10^{-10}$ & $7.5 \times 10^{-10}$ 
& $6.9 \times 10^{-10}$ \\
unabsorbed flux\tablenotemark{d} & 0.01--100 keV (ergs cm$^{-2}$ sec$^{-1}$) & $5.7 \times 10^{-9}$ & 
$4.8 \times 10^{-9}$ & $4.4 \times 10^{-9}$ \\ 
& $\chi^2/{\rm d.o.f.}$ & $1347/1155$ & $984/991$ & $995/901$
\enddata
\tablenotetext{a}{The electron temperature is first estimated only from the GSO data 
and it is fixed at the best-fit value in the joint fit with the XIS and HXD spectra. 
Although the upper limit for Epoch-1 is not constrained, we adopted the 
best-estimate value, $kT_{\rm e} = 57$ keV, in the simultaneous fit.}
\tablenotetext{b}{The emissivity index. The radial dependence of emissivity ($\epsilon$) 
is expressed with $\epsilon \propto R^{-\alpha}$.}
\tablenotetext{c}{The normalization of the Gaussian is linked to the solid angle of 
{\tt ireflect} component so that the equivalent width with respect to the reflection 
continuum is always 1.0 keV.}
\tablenotetext{d}{The flux is corrected for dust-scattering.}
\end{deluxetable*}

It has been suggested that the Comptonization cloud in the 
low/hard state has a complex structure. Using {\it Suzaku} broad-band 
X-ray data, \citet{tak08, mak08, shi11b, yam13} found that the 
time-averaged spectra can be described with two Comptonization 
components that have different optical depths.
Moreover, \citet{yam13} detected the second Comptonization component by
analyzing the spectral variability. To test the ``double Compton'' model
in the H 1743$-$322 data, we add one more {\tt nthcomp} component and
its accompanying reflection component to our final model, and fit the
time-averaged spectra. We find, however, that the inclusion of the
second Comptonization component does not improve the quality of fit.

\subsection{Analysis of Short-term Spectral Variability}

We often find it difficult to accurately estimate the cool disk
component in the low/hard state by modeling time-averaged spectra,
because it is buried in the dominant Comptonized component and coupled
with the other structures such as the interstellar absorption and dust
scattering. In these cases, the analysis of the short-term spectral
variability can be a more powerful approach to separate the disk
component. The standard disk and Comptonization in the corona have
different properties of spectral variability on the $\sim$1-sec
timescale; the former generally shows significant variation, while 
the latter is more stable \citep[e.g.,][]{chu01}. 

\begin{figure}
\plotone{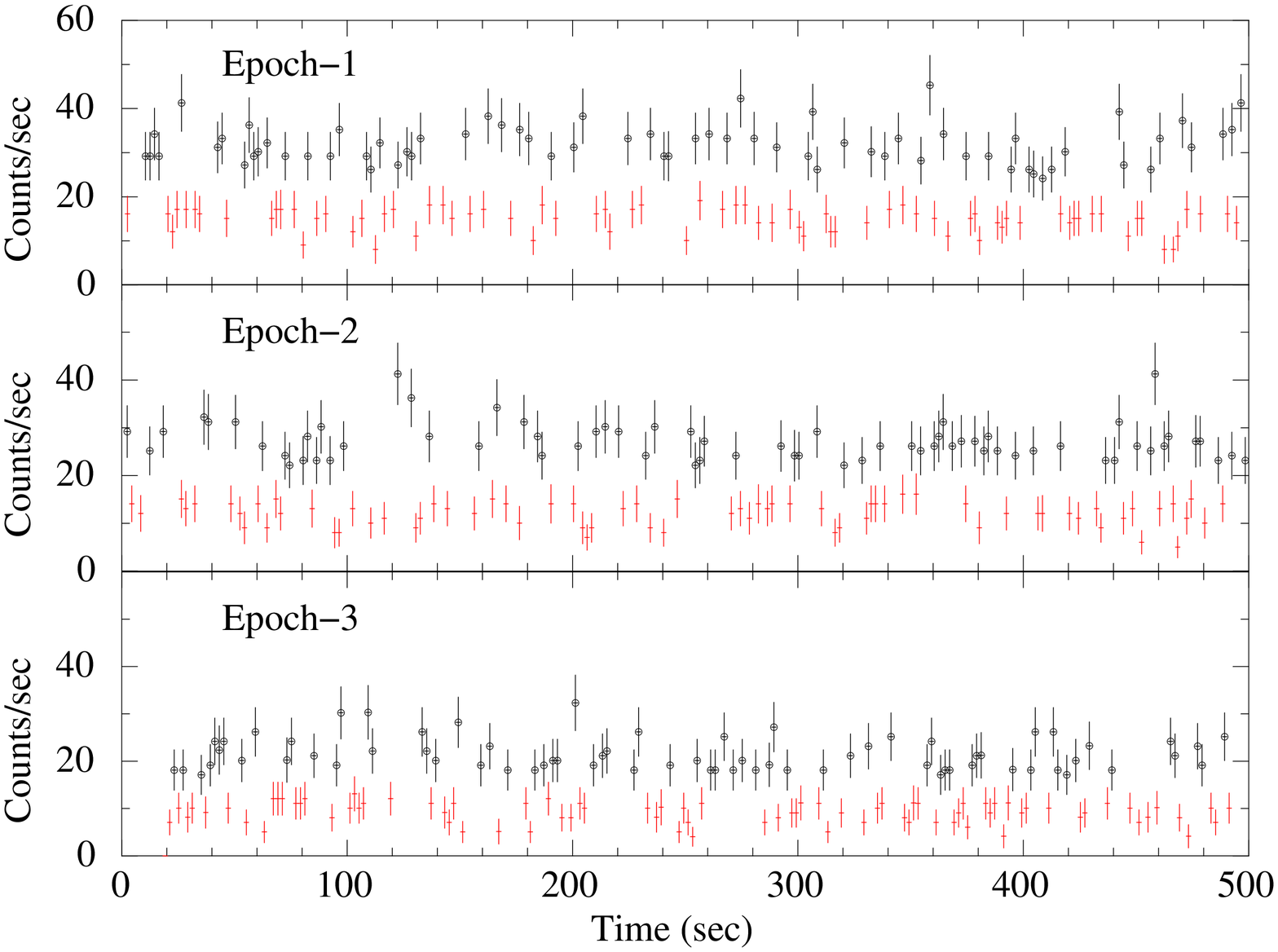}
\caption{The XIS-1 light curves of Epoch-1 (top), Epoch-2 (middle), 
and Epoch-3 (bottom) in 1--10 keV for the high- (black) and  
low-intensity (dark gray) phases. The bin size is 2.0 sec.\label{fig_lc_isort}}
\end{figure}

Here we apply ''intensity-sorted spectroscopy'' described in \citet{mak08} and 
\citet{yam13} to our {\it Suzaku} data. Using this technique, \citet{yam13} 
successfully separated the cool disk component from the highly variable 
Comptonization in Cyg X-1. We define the high- and low-intensity phases in 
the same manner as theirs; 
\begin{equation}
\{ t| C(t) > (1+f) \overline{C(t)_T} \}
\end{equation}
and
\begin{equation}
\{ t| C(t) < (1-f) \overline{C(t)_T} \},
\end{equation}
respectively, where $C(t)$ is the count rate at the time $t$ for an XIS 
detector and $\overline{C(t)_T}$ is that averaged over $[t - T/2, t + T/2)$. 
According to these criteria, we define the time intervals of high- and 
low-intensity phases using XIS-1 light curves in the 1--10 keV band in
2.0 sec binning. We set $f = 0.2$ and $T = 64$ sec, so that we obtain
both the good photon statistics and a sufficiently high contrast of the
intensity between the two phases. The XIS-0$+$XIS-3, PIN, and GSO
spectra in these two phases are then extracted from the intervals
determined by XIS-1. 
Thus, by using the independent dataset for the definition of the 
time regions, we avoid selection biases merely caused by statistical 
fluctuation in the photon counts in making the intensity-sorted spectra. 
A summary of the intensity selection is given in Table~\ref{tab_isort_log}, 
and the XIS-1 light curves in low- and high-intensity phases are presented
in Figure~\ref{fig_lc_isort}.

The high and low-intensity spectra ($H(E)$ and $L(E)$, respectively) can be 
separated to the constant component ($d(E)$) and variable components ($h(E)$ 
and $l(E)$) as 
\begin{equation}
H(E) = \omega(E) (d(E)+h(E)) \label{eq_h}
\end{equation}
and
\begin{equation}
L(E) = \omega(E) (d(E)+l(E)), \label{eq_l}
\end{equation}
where $\omega(E)$ represents the photo-electric absorption (which can be
regarded as constant on the $\sim$1-sec timescale). Unlike \citet{yam13}, 
the constant component $d(E)$ in our data contains not only the 
direct disk emission but also the scattered-in component, 
whose short-term variability is smoothed out 
due to the difference of the light traveling time of each scattered 
photon. On the other hand, the observed variable component is composed 
of the Comptonized photons without experiencing dust-scattering. For 
simplicity, we assume the best-fit models obtained from the time-averaged 
spectra as the scattered-in disk and Comptonization components, and subtract 
them from both the high- and low-intensity spectra before calculating the 
spectral ratio $H(E)/L(E)$. 

\begin{figure}
\plotone{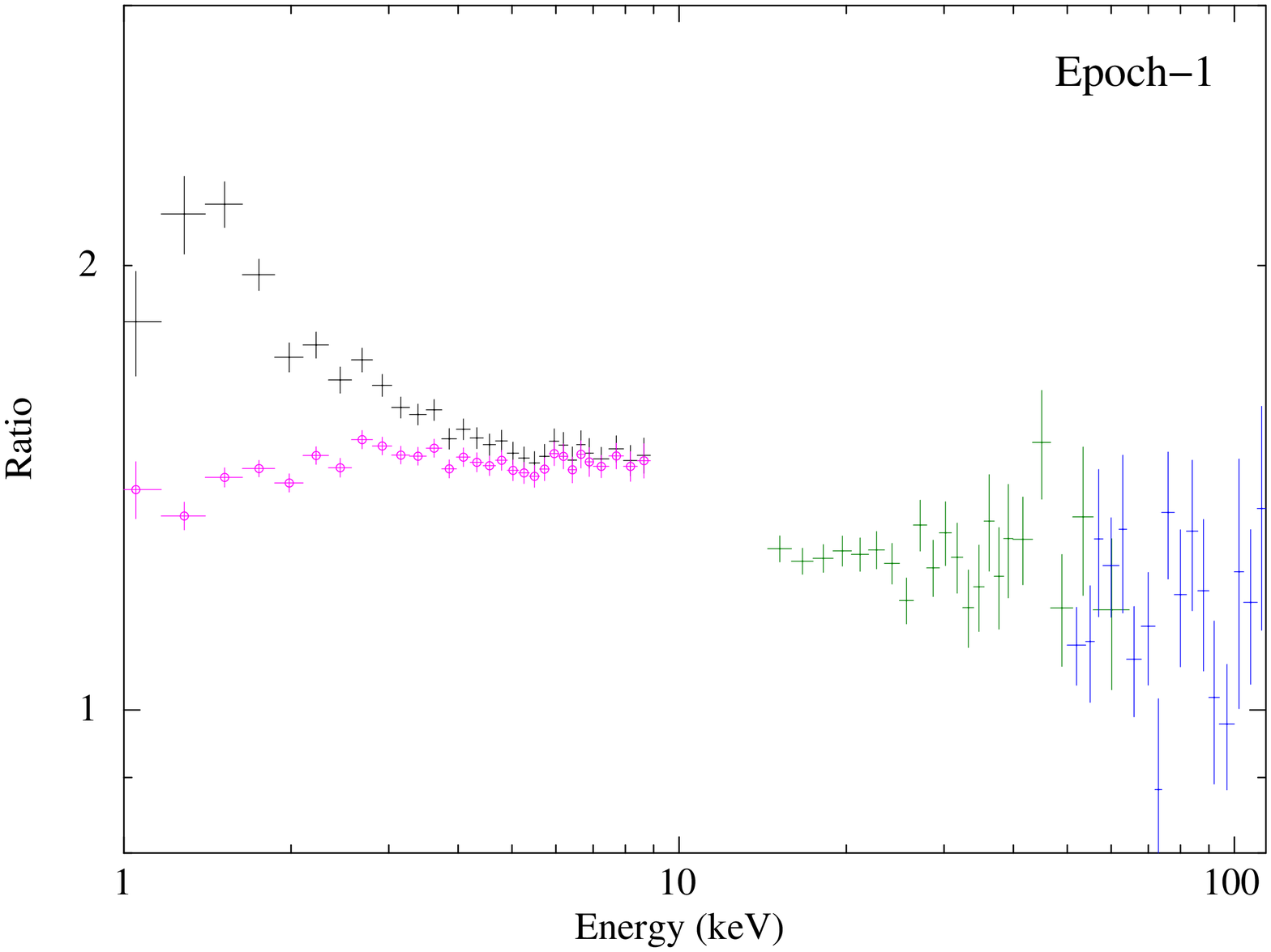}
\plotone{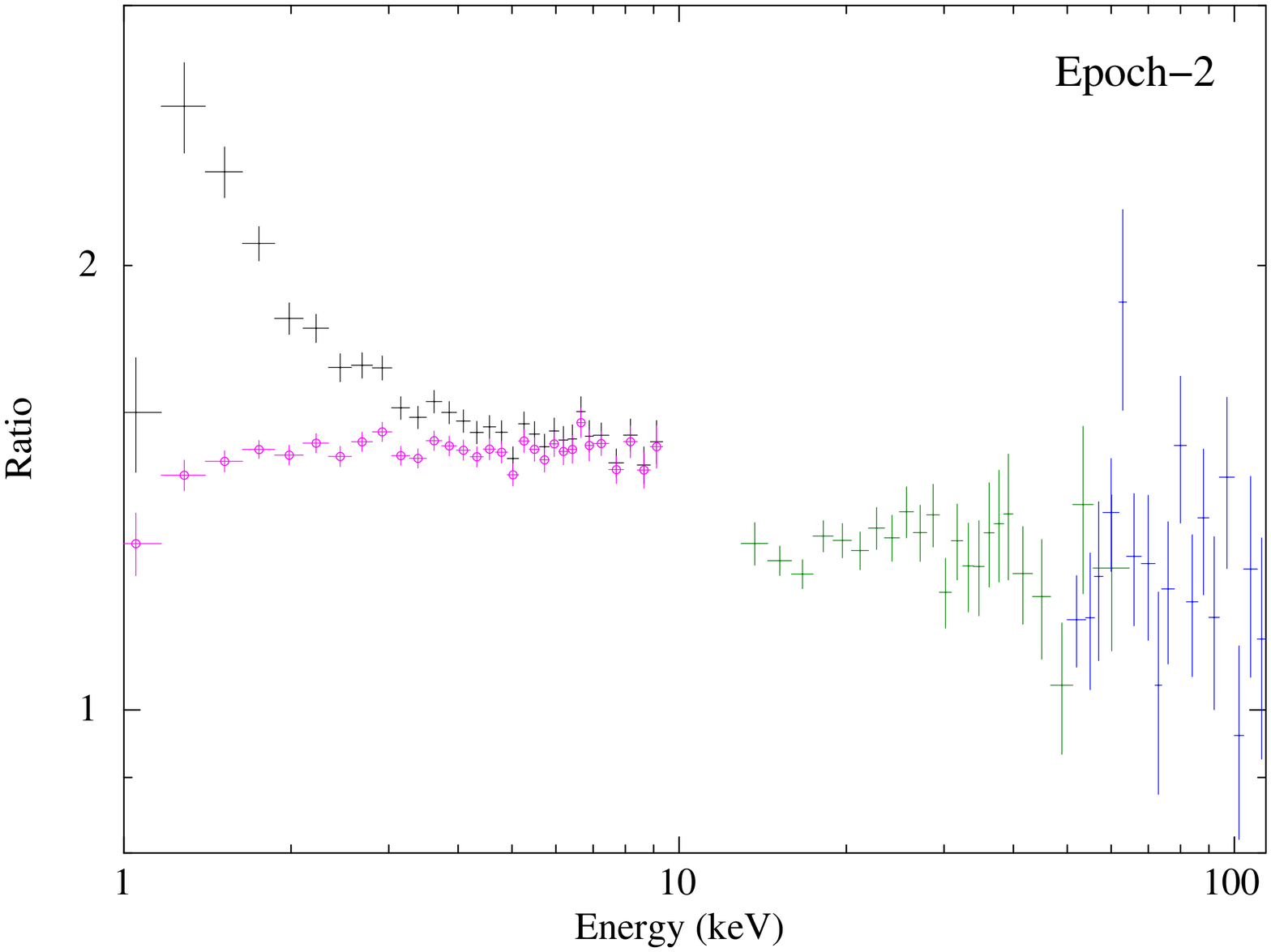}
\plotone{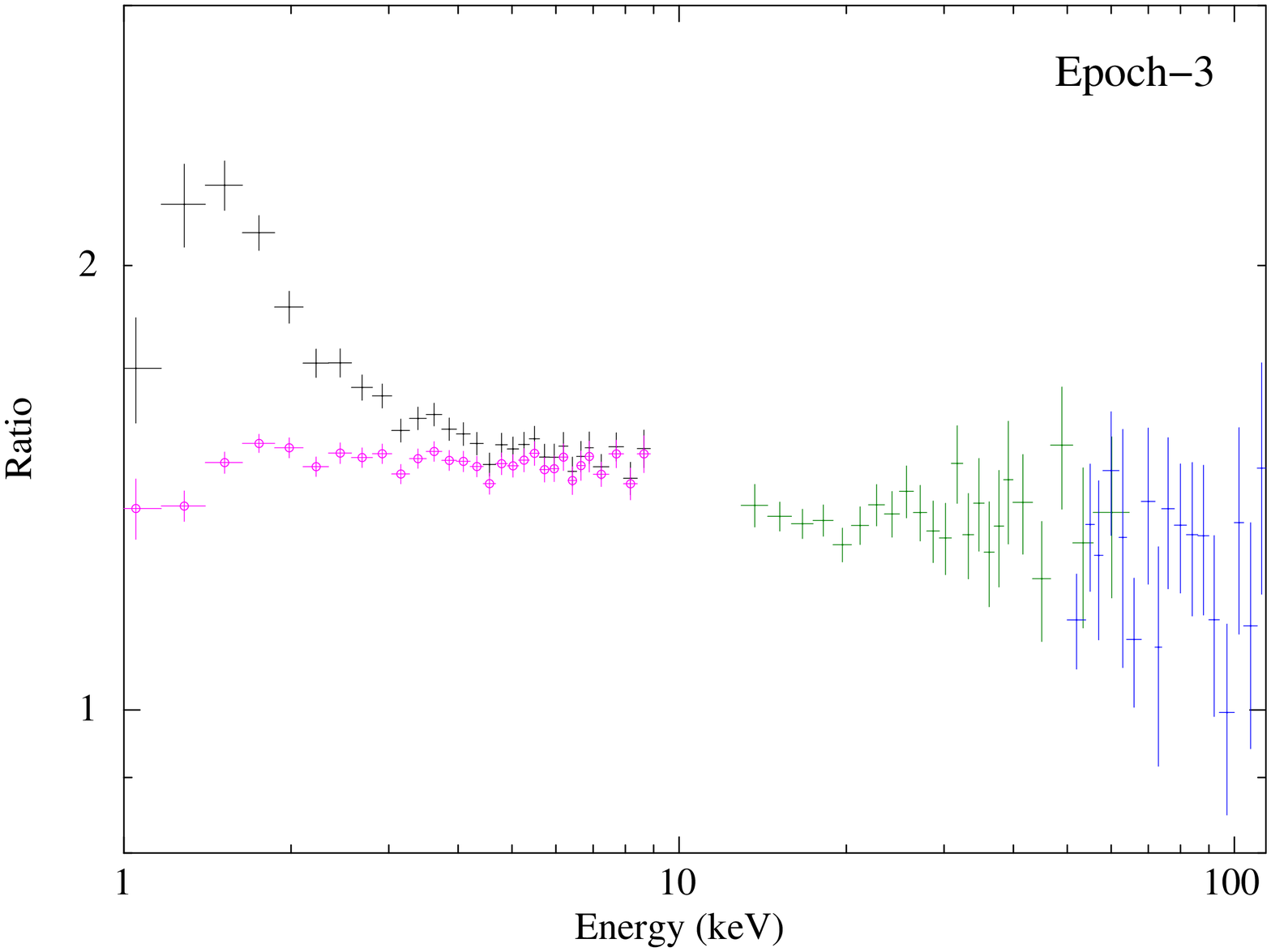}
\caption{The ratios between the high-intensity and low-intensity spectra in 
Epoch-1 (top), Epoch-2 (middle), and Epoch-3 (bottom). The ordinate axis in 
each panel is logarithmically scaled. 
The scattered-in components are already subtracted from the XIS spectra.
The same colors as Figure~\ref{fig_specfit} are used. The XIS-1 data, 
which are used to determine the intervals of high- and low-intensity phases, 
are not shown here, because they are affected by Poisson fluctuation of signal 
counts. The XIS spectral ratio including the scattered-in component is also 
plotted in pink (open circle). 
\label{fig_spec_ratio}}
\end{figure}

Figure~\ref{fig_spec_ratio} plots the resultant spectral ratio in each  
epoch. The overall shape is very similar to that of the Cyg X-1 spectra 
in the low/hard state obtained in \citet{yam13}. A turnover below 
$\sim 2$ keV can be seen, suggesting a significant suppression of 
variability due to the contribution of constant disk component at low 
energies. The spectral ratios of 
XIS-0$+$XIS-3 without removing the scattered-in component are also 
presented in pink for comparison. These demonstrate how significantly 
dust scattering affect the results. We note that the turnover is intrinsic, 
not generated by the dust scattering. The scattered-in component does not 
have such a steep turnover, and the decline of variability is seen 
in the spectral ratio including the scattered-in component as well.

Following the procedure of \citet{yam13}, we assume that the ratio 
$H(E)/L(E)$ can be expressed with a single power-law ($\alpha E^{\beta}$) 
in $E>2$ keV. We determine $\alpha$ and $\beta$ by fitting the data 
in the 2--4 keV band (above which they may affected by the reflection 
component). Using these parameters and Equation~\ref{eq_h} and \ref{eq_l}, 
we derive
\begin{equation}
\omega(E) d(E) = \frac{\alpha E^{\beta} L(E) - H(E)}{\alpha E^{\beta} - 1},
\end{equation}
where $d(E)$ is the direct disk spectra. 
$d(E)$ is then fitted with {\tt phabs*Dscat*diskbb} to estimate the inner 
temperature and the flux of the constant disk emission. Here the scattering 
fraction of the {\tt Dscat} model is fixed at 0.0 (i.e., only the effect 
of scattering-out is considered).
We assume $N_{\rm H} = 2.0 \times 10^{22}$ cm$^{-2}$, which is the averaged 
value of those determined by the time-averaged spectra in the individual 
epochs. However, the parameters are only very weakly constrained due to 
the poor statistics in the 
soft X-ray band around the turnover of the spectral ratios, although the values 
are consistent with those estimated from the time-averaged spectra. 
The fitting parameters are summarized in Table~\ref{tab_isort}.

\begin{deluxetable}{lcccccc}
\tablecaption{Summary of the high- and low-intensity selection. \label{tab_isort_log}}
\tablewidth{0pt}
\tablehead{
\colhead{} & \multicolumn{2}{c}{Epoch-1} 
& \multicolumn{2}{c}{Epoch-2} & \multicolumn{2}{c}{Epoch-3} \\
& \colhead{High} & \colhead{Low} & \colhead{High} & \colhead{Low} & \colhead{High} & \colhead{Low} 
}
\startdata
frac. exp. (\%)\tablenotemark{a} & 24 & 27 & 24 & 28 & 25 & 29 \\
ave. rate (cnt sec$^{-1}$)\tablenotemark{b} & 
27 & 18 & 22 & 15 & 20 & 14
\enddata
\tablenotetext{a}{The fraction of time that the source was in high- 
and low-intensity phases with respect to the total exposure for each epoch.}
\tablenotetext{b}{Averaged XIS-0 count rates in 1--10 keV.}
\end{deluxetable}

\begin{deluxetable}{lccc}
\tablecaption{Results from the analysis of intensity-sorted spectra. \label{tab_isort}}
\tablewidth{0pt}
\tablehead{
\colhead{Parameter} & \colhead{Epoch-1} & \colhead{Epoch-2} 
& \colhead{Epoch-3} 
}
\startdata
\mbox{\boldmath $ h(E)/l(E)$}\tablenotemark{a} & & & \\
$\alpha$ & $2.1 \pm 0.1$ & $2.1 \pm 0.2$ & $2.0 \pm 0.1$ \\
$\beta$ & $-0.24 \pm 0.06$ & $-0.23 \pm 0.06$ & $-0.19 \pm 0.06$ \\ 
$\chi^2/{\rm d.o.f.}$ & $6.1/6$ & $8.1/6$ & $13.4/13$ \\ \hline 
 \mbox{\boldmath $\omega(E) d(E)$}\tablenotemark{b} & & & \\
$N_{\rm H}$ ($10^{22}$ cm$^{-2}$) & 2.0 (fixed) & 2.0 (fixed) & 2.0 (fixed) \\
scatf\tablenotemark{c} & 0.0 (fixed) & 0.0 (fixed) & 0.0 (fixed) \\
$T_{\rm in}$ & $0.22^{+0.40}_{-0.08}$ & $<0.33$ & $0.11^{+0.19}_{-0.06}$ \\
norm & $< 8 \times 10^{5}$ & $> 2 \times 10^3 $ & $< 1 \times 10^{12}$ \\
$\chi^2/{\rm d.o.f.}$ & $3.1/6$ & $4.2/3$ & $11.3/8$ 
\enddata
\tablenotetext{a}{$h(E)/l(E)$ is fitted with $\alpha E^{\beta}$ in the 2--4 keV band.}
\tablenotetext{b}{$\omega(E) d(E)$, calculated by using the best-fit values of 
$\alpha$ and $\beta$, is fitted with {\tt phabs*Dscat*diskbb}.}
\tablenotetext{c}{The scattering fraction in the {\tt Dscat} model.}
\end{deluxetable}

\section{Near-infrared and optical observations and the results}

In addition to the X-ray observations with {\it Suzaku}, we performed 
infrared photometric observations of H~1743$-$322 in the $J$  (1.25 $\mu$m), 
$H$ (1.63 $\mu$m), and $K_{\rm s}$ (2.14 $\mu$m) bands, with the SIRIUS 
camera \citep{nag03} on the 1.4m {\it IRSF} telescope 
at South African Astronomical Observatory (SAAO)
during the 2012 outburst. Optical observations through $R_{\rm C}$ and
$I_{\rm C}$ filters were also carried out by using HOWPol
\citep[Hiroshima One-shot Wide-field Polarimeter;][]{kaw08} attached to
the 1.5m {\it Kanata} telescope at Higashi-Hiroshima Observatory.  The
counterpart of H 1743$-$322 was not detected in any bands, however.  The
typical seeing was $\approx$ 1''.8 for the {\it IRSF} $J$ band 
and 2''.3 for the {\it Kanata} $R_{\rm C}$ band in full width at
half maximum.

In the analysis of the {\it IRSF} data, we combine all the frames taken
in one night and perform dark-subtraction, flat-fielding,
sky-subtraction, and combining the dithered images using the SIRIUS
pipeline software running on IRAF \citep[Image Reduction and Analysis
Facilities;][]{tod86} version 2.16. H 1743$-$322 is located in a
crowded region close to the Galactic center, and the near-infrared
counterpart in the {\it IRSF} images is buried in nearby bright
sources. We estimate the upper flux in each band from the counts on the
pixel at the position of H 1743$-$322.  Count-flux conversion is made by
comparison with the stellar photometry of nearby sources with the 2MASS
magnitude \citep{skr06}. We reduce the {\it Kanata} data in a standard
way for the aperture photometry of optical CCD data. The upper
magnitudes in $R_{\rm C}$ and $I_{\rm C}$ band are determined by the
detection limits of a point source in the image, estimated from the
background level of source-free regions near H 1743$-$322.

\begin{figure}[tbp]
\plotone{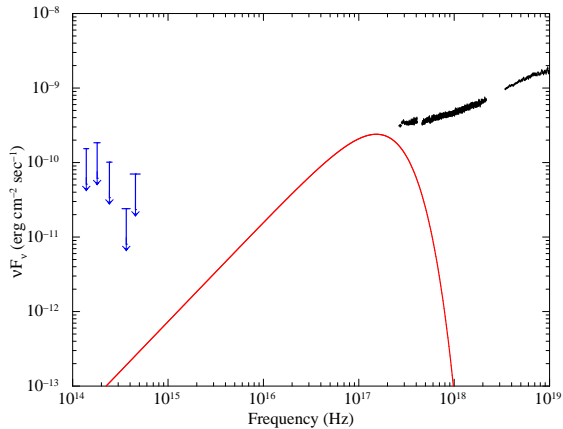}
\caption{The spectral energy distribution of H 1743$-$322 on 2012 October 12. 
The upper limits in the extinction-corrected 
optical ($R_{\rm C}$ and $I_{\rm C}$) and near-infrared 
($J$, $H$, and $K_{\rm S}$ bands) fluxes 
are plotted in blue arrows. The black points 
are the {\it Suzaku} spectra, corrected for neutral absorption and dust 
scattering. Red solid line shows the intrinsic disk emission including the 
Comptonized photons, where the outer disk is assumed to extend to infinity. 
\label{fig_SED}}
\end{figure}

Figure~\ref{fig_SED} presents the upper limits of the optical and 
near-infrared fluxes on 2012 October 12, together with the 
contemporaneous {\it Suzaku} spectrum (in Epoch-3). 
The flux limits in the optical and near-infrared bands are corrected 
for the Galactic extinction. Substituting the hydrogen column density 
of H 1743$-$322 ($N_{\rm H} \approx 2 \times 10^{22}$ cm$^{-2}$) into the 
$N_{\rm H}$ versus $A_{\rm V}$ relation 
\citep[where $A_{\rm V}$ represents the extinction in the $V$ band;][]{pre95}, 
we derive $A_{\rm V} \approx 11$. We convert $A_{\rm V}$ 
to the extinction in the $R_{\rm C}$, $I_{\rm C}$, $J$, $H$, and 
$K_{\rm S}$ bands using the extinction curve for $R_{\rm V} = 3.1$
given by \citet{car89}.
The upper limits of the extinction-corrected absolute 
$R_{\rm C}$, $I_{\rm C}$, $J$, $H$, and $K_{\rm S}$ magnitudes 
are estimated to be $-1.9$, $-0.5$, $-3.8$, $-5.2$, and $-5.8$, 
respectively.
We find that if the companion is a main sequence star, 
it should be a late-B or a less massive star \citep{wai92}.

The upper limits are consistent with the intrinsic disk flux including 
the Compton-scattered photons (shown in red in Fig.~\ref{fig_SED}), 
and with the possible contribution of the synchrotron emission from 
the compact jet. If the positive relation between radio and X-ray fluxes 
found by \citet{cor11} holds in the 2012 outburst, the {\it Suzaku} flux 
in Epoch-3 ($5.2 \times 10^{-10}$ erg cm$^{-2}$ 
sec$^{-1}$in the 3--9 keV band) corresponds to 0.68 mJy at 8.5 GHz flux. 
Assuming that the flat optically-thick synchrotron spectrum 
extends up to the optical bands, we estimate the jet contribution to be 
$\nu F_\nu \approx 4 \times 10^{-12}$ erg cm$^{-2}$ sec$^{-1}$ at 
$5 \times 10^{14}$ Hz.
This is more than $\approx$10 times lower than the optical and near-infrared 
upper flux limits.

\section{Discussion}

\subsection{Failed Outburst in 2012 October}

We observed H 1743$-$322 with {\it Suzaku} at the flux peak and early
decaying phase during the outburst in 2012 September and October. The
time-averaged spectra in all the three epochs are approximated by
a power-law with a photon index of $\approx$1.6, which is 
well within the typical values in the low/hard state 
\citep[1.5--2.0; e.g.,][]{don07}. The PIN power spectra in 
the 14--70 keV band show strong variability above $\approx$0.1 Hz with a 
power of $1 \times 10^{-2}$ (rms$^2$/Hz$^2$), which most probably 
corresponds to the flat part of the band-limited noise. These X-ray 
spectral and timing properties support that the source was in the 
low/hard state during the {\it Suzaku} observations.

As noticed in Fig.~\ref{fig_longtermLC}, the long-term MAXI data do 
not show significant softening around the peak flux. This suggests 
that the state transition from the low/hard state to the high/soft 
state (hard-to-soft transition) did not occur in the late 2012 outburst. 
This type of outburst is often classified as a ``failed outburst'', which 
exhibits no hard-to-soft transition or an incomplete transition only 
reaching the intermediate state.
A few failed outbursts have been reported in H 1743$-$322 so far 
\citep{cap09,che10}. In addition to this source, there are few 
other BHXBs that underwent failed outbursts, like XTE J1550$-$564 
and MAXI J1836$-$164 \citep{stu05,fer12}. 
The 1--500 keV unabsorbed luminosity in Epoch-1 (calculated 
from the best-fit model) is estimated to be $6.1 \times 10^{37}$ 
$D_{8.5}^2$ erg sec$^{-1}$.
This is a few times lower than the luminosity at which H 1743$-$322 
generally makes a hard-to-soft transition 
\citep[e.g.,][]{mcc09,zho13}, suggesting that the mass 
accretion rate was not enough to trigger a transition.

\subsection{Structure of Inner Disk and Corona}

The time-averaged {\it Suzaku} spectra are well described with thermal
emission from the standard disk, its Comptonized emission by a hot 
corona, and the reflection component from the accretion
disk. The overall spectral shape is not largely changed in the three epochs, 
while the 2--20 keV fluxes in Epoch-2 and Epoch-3 are
$\approx$20\% lower than that in Epoch-1. Although the inner disk
temperature and the photon index of the Comptonization are decreased by
10--30\% and 0.03 from Epoch-1 to the latter two epochs, the other
parameters remain the same within their 90\% confidence ranges. We
find that the double Comptonization model is not necessary to reproduce
the spectra. The same was suggested with the {\it Suzaku} spectrum of
MAXI J1305$-$704 in the low/hard state at $\sim 0.01 L_{\rm Edd}$ 
\citep{shi13}. 
In the case of Cyg X-1, the softer Comptonization component is more clearly 
seen in the brighter low/hard (or hard intermediate) states \citep{yam13}. 
Thus, the luminosities of H 1743$-$322 in {\it Suzaku} observations 
($\approx 0.04$ $D_{\rm 8.5}^2 M_{\rm 10}^{-1}$ $L_{\rm Edd}$ in the 1--500 keV 
band) may not be sufficiently high to detect the softer Comptonization 
component significantly.

As noticed in Fig.~\ref{fig_spec}, the GSO spectra suggest  
the evidence of the high-energy rollover around 50--100 keV, likely 
correspond to the electron temperature of the Comptonized corona.
From the spectral fit with the GSO data, the electron temperature 
is estimated to be $kT_{\rm e} \approx 60$ keV.
We note, however, that the value may be somewhat overestimated
because the {\tt nthcomp} model does not include relativistic 
effects properly; its output spectrum drops more sharply above 
the rollover than the that of the real Comptonization spectrum 
\citep[see][]{don10}. 
Previous observations of H 1743$-$322 with {\it INTEGRAL} and 
{\it Suzaku} at similar luminosities reported the detection of 
the cutoff energy \citep[e.g.,][]{cap05, blu10}. 
The anti-correlation between the electron temperature and the 
X-ray luminosity in the bright ($\lesssim 0.1 L_{\rm Edd}$) 
low/hard state was suggested by \citet{miy08} with the {\it RXTE} 
\citep[Rossi X-ray Timing Explorer;][]{bra93} spectra of GX 339$-$4.
\citet{chi10} also found in the {\it RXTE} data of Swift J1753.5$-$0127 
that the electron temperature is relatively low in the bright low/hard 
state like in our case, and that it moves to higher energies in 
accordance with the decline of an outburst. 
It would be explained if inverse Compton cooling in the corona is 
more efficient by a relatively higher input rate of seed photons
in brighter periods \citep{miy08}.

We find that the X-ray spectra are dominated by the Comptonized 
component and that the direct disk emission is very weak in the 
{\it Suzaku} bandpass. The disk temperature estimated from the 
time-averaged spectra is much smaller than typical values in
the high/soft state \citep[$\approx$1 keV; e.g.,][]{mcc06}. 
We calculate the inner disk radius from the intrinsic disk flux 
including the disk photons consumed by Compton-scattering in the 
corona in each epoch. The radii are found to be 1.3--2.3 times larger
than that in the high/soft state, supporting the idea that the standard
disk does not extend to the ISCO in the low/hard state. This is 
consistent with the somewhat smaller solid angle of the reflection 
component than those obtained in the very high state 
by using similar spectral models \citep{tam12, hor14}.

We note, however, the calculated inner radii should be 
taken with caution, since this is somewhat dependent on spectral 
modeling. We considered only the statistical uncertainties of 
the inner disk temperatures, but additional errors may be 
produced in modeling the effects of dust scattering.
Also, the spectral hardening factor of disk emission for 
deriving the actual inner radii could be different between 
the high/soft and low/hard states \citep[e.g.,][]{shi95}. 
Moreover, we assumed that all the seed photons are originated 
from the disk, but if other emission components contribute, 
the inner radii could become smaller. 
For more precise estimation of the inner disk radii, we 
have to assess these possible uncertainties thoroughly 
by using higher-quality wide-band spectra, particularly 
from the sources with a much smaller absorption column. 
This is left for future work.

We successfully detect the weak, cool disk component independently of
the time-averaged spectral modeling, using the spectral variability on a
short ($\approx$1 sec) timescale. The overall profile of spectral ratio
between high- and low-intensity phases is very similar to what
\citet{yam13} found in Cyg X-1 data during the low/hard state. The power
of short-term variability clearly declines below 1--2 keV, suggesting
that the constant standard disk component contributes to the soft X-ray
flux.  Modeling the profile of the spectral ratio in the same manner as
\citet{yam13}, we estimate the inner radius and temperature of the disk
component. Although the resultant parameters have large uncertainties
due to poor statistics, they are consistent with those obtained from the
time-averaged spectra.

The mass accretion rate of the disk can be derived through 
$L_{\rm disk} = GM_{\rm BH} \dot{m}_{\rm disk}/(2R_{\rm in})$, where 
$L_{\rm disk}$ and $\dot{m}_{\rm disk}$ represent the luminosity and 
mass accretion rate of the disk component, respectively. We estimate 
$L_{\rm disk} = 2.5 \times 10^{36}$ $D_{8.5}^2$ erg sec$^{-1}$ (the 
unabsorbed disk luminosity in the 0.01--500 keV band) and 
$R_{\rm in} = 120$ $D_{8.5} (\cos i/\cos 75^\circ)^{-1/2}$ km (for 
Epoch-1) and thus we have $\dot{m}_{\rm disk} = 4.5 \times 10^{16}$ 
$D_{8.5}^3$ $M_{10}^{-1} (\cos i/\cos 75^\circ)^{-1/2}$ g s$^{-1}$. 
The accretion rate of corona can be calculated as $L_{\rm c} 
= \eta \dot{m}_{\rm c} c^2$, where $\eta$, $L_{\rm c}$, and 
$\dot{m}_{\rm c}$ are the radiative efficiency, 
the luminosity and the mass accretion rate of the corona, respectively. 
Using the best-fit {\tt nthcomp} component for Epoch-1 as $L_{\rm c}$ 
($5.1 \times 10^{37}$ $D_{8.5}^2$ erg sec$^{-1}$), 
we obtain $\dot{m}_{\rm c} = 5.7 \times 10^{17}$ $D_{8.5}$ g s$^{-1}$ 
if the corona is radiatively efficient ($\eta = 0.1$ is assumed).  
It can be much more larger in the case of a radiatively inefficient 
flow. Thus, $\dot{m}_{\rm c}$ is much larger than 
$\dot{m}_{\rm disk}$, regardless of the radiative efficiency. This is 
inconsistent with the simple truncation disk model, in which the 
standard disk makes transition at the truncated radius into the 
hot inner flow with the same mass accretion rate.

A possible explanation is that there may be a separate coronal 
accretion flow with a large scale height extending from the outer 
edge of the disk to the black hole, and that its mass accretion rate 
is much larger than that of the thin disk. 
There is another possibility, which reconciles the mass accretion 
rates of the disk and corona. If the soft X-ray component is not the 
disk itself, but is instead originated from a much smaller, hot clumps 
torn off the inner edge of a much larger disk \citep{chi10, yam13}, the 
true disk component has larger normalization but is truncated at a 
larger radius. In this case, $\dot{m}_{\rm disk}$ can be comparable to 
$\dot{m}_{\rm c}$, and the true inner disk temperature can be much lower, 
making impossible to detect the disk component in the soft X-ray bandpass.
The cool disk component peaking in the EUV band was actually detected in a 
low/hard state spectrum of XTE J1118$+$480 \citep{esi01}, along 
with a much smaller soft X-ray component \citep{rei09}, which would 
be from the clumps \citep{chi10}.

\subsection{Ionized Absorber and Disk Wind}

\citet{mil06} discovered blue-shifted, ionized absorption lines at 6.7
keV and 7.0 keV (corresponding to He-like and H-like iron K$\alpha$
lines, respectively) in the Chandra high-resolution spectra of
H~1743$-$322 in the high/soft and very high (or soft intermediate)
states. These lines, often seen in high inclination sources
\citep[e.g.,][]{ued98, kot00, lee02}, are likely to be originated from
the disk wind. They show ``state-dependent'' behavior and are rarely
detected in the low/hard state \citep[e.g.,][]{nei09,pon12}. For H
1743$-$322, non detection of the iron absorption lines in the low/hard
state was previously reported by \citet{blu10} and \citet{mil12}.

Our {\it Suzaku} data do not exhibit any significant absorption features 
apparently. To check the presence of the ionized iron absorption lines, 
we add two negative Gaussian components to the best-fit model for the 
time-averaged spectra. Here we assume a wind velocity of 
300 km sec$^{-1}$ and a 1-$\sigma$ line width of 20 eV (which can be 
regarded as representative values of \citealt{mil06}) and fix all the 
parameter except for the normalizations of continuum and line 
components. We estimate the 90\% upper limits of the equivalent widths 
in Epoch-2 to be 22~eV for both H-like and He-like iron lines, which 
are similar or slightly larger values than those obtained in \citet{mil06}. 
Thus, we cannot exclude the presence or disappearance of the ionized 
iron absorption lines.

\subsection{Origin of QPO}

A weak low-frequency (LF) QPO is detected at 0.1--0.2~Hz in each
epoch. We find that the QPO frequency becomes $\approx$30\% lower in
accordance with the $\approx$20\% decrease of the hard X-ray luminosity
in the 15--50~keV band.
LF QPOs (below 10~Hz) were detected in many BHXBs during the low/hard
and intermediate states, such as XTE J150$-$564
\citep[e.g.,][]{cui99,sob99} and GX 339$-$4
\citep[e.g.,][]{miy91,mot11,tam12}. Previous studies suggested that
their frequencies are positively correlated with the photon index of the
Comptonization below $\approx$10~Hz, at which the correlation saturates
\citep{vig03,tit04,tit05}. Our results follow this ubiquitous
correlation between the QPO frequency and the photon index; the photon
index becomes by $\approx$0.03
lower in the $\approx$30\% decrease of the QPO frequency.

The correlations of the observed LF-QPO frequencies with the X-ray flux
and the photon index lead us to invoke the idea that they are associated
with the inner disk structure. \citet{ing09} suggested the possibility that 
the LF QPOs are originated in the Lense-Thirring precession of the hot 
inner flow extended between the black hole and the inner edge of the
disk. Their model can describe both the QPO frequencies and the 
observed X-ray spectrum, and predicts the anti-correlation with the
inner disk radius and the QPO frequency. 
In the truncation disk model, the inner radius recedes as the mass 
accretion rate is decreased \citep[see][]{don07}. At the same time, 
the power of seed photons illuminating the hot flow becomes weaker, 
making the observed X-ray spectrum harder. The decrease of the LF-QPO 
frequency can be explained if it reflects the timescale at the inner 
radius, where the standard disk is replaced to the hot flow \citep{ing09}. 
\citet{axe14} extracted the spectrum of the pure QPO component from 
the RXTE data of XTE J1550$-$564 and found that it can indeed be 
interpreted by the Comptonization in the hot inner flow. Although the 
inner disk radii that we estimated from the {\it Suzaku} data in each 
epoch have large uncertainties, they are compatible to the 
anti-correlation trend between the QPO frequency and the inner disk.

The QPO frequencies correspond to 40--50 $R_{\rm g}$ in the
Lense-Thirring precession model of \citet{ing09}, if a black hole 
mass of $10 M_\sun$ is assumed. This is about several to ten times 
larger than what we estimated from the time-averaged spectra. 
The possible reason of this inconsistency would be that the QPO 
was generated by a more complicated mechanism, or that we are 
underestimating the true inner disk radius, as is also suggested 
in Section 6.2 by assuming that the mass accretion rate through 
the disk and that in the coronal flow are the same.
A separate, small variable soft component could simultaneously 
allow the QPO to be consistent 
with a Lense-Thirring origin, and for the corona to be fed directly 
by the mass accreting through the disk.

\section{Summary and Conclusion}

We observed the black hole transient H 1743$-$322 
with {\it Suzaku} on three occasions during the outburst in 2010 
October, which provided one of the best quality broad band X-ray 
spectra of this source in the low/hard state. 
The results are summarized as follows:

\begin{enumerate}

\item We find that the observed X-ray spectra are significantly affected
by dust scattering in the interstellar medium with $N_{\rm H} \approx 2
\times 10^{22}$ cm$^{-2}$, which must be corrected for accurate spectral
analyses.

\item The time-averaged spectra are dominated by a strong Comptonization 
in the corona with an electron temperature of $\approx$60~keV.
The disk reflection component with $\Omega/2\pi \approx 0.6$ is 
detected. The estimated inner disk radius is larger than that 
in the high/soft state, suggesting that the standard disk is truncated
at 1.3--2.3 times larger radii than the ISCO during the {\it Suzaku}
observations. 

\item We estimate the cool disk component by investigating short-term
spectral variability on the $\sim$1-sec timescale, independently of the
time-averaged spectral modeling. We find that the spectral ratio between
high- and low-intensity phases has a very similar shape to those of Cyg
X-1 in the low/hard state, indicating the presence of stable disk
emission at low energies.

\item A weak low-frequency QPO was detected at 0.1--0.2~Hz. We find that
the QPO frequency becomes lower as the X-ray luminosity and photon index
decrease. This may be explained by the evolution of the disk truncation 
radius.

\end{enumerate}

\acknowledgments

We appreciate the {\it Suzaku} operation team for arranging and carrying out 
the ToO observations.  We thank Takayuki Yuasa for his help in examining the 
contamination of Galactic ridge X-ray emission. We are grateful to the 
{\it Kanata} team, Hiroshima University, for performing the optical 
observations. We also thank 
Kohji Tsumura, Mai Shirahata, Sudhanshu Barway, Daisuke Suzuki, and 
Fumio Abe for carrying out the {\it IRSF} observations. We 
appreciate Tatsuhito Yoshikawa for his help in analysing the {\it IRSF} data.
This research has made use of the MAXI data provided by RIKEN, JAXA and 
the MAXI team. We utilized data products from the Two Micron All Sky Survey, 
which is a joint project of the University of Massachusetts and the Infrared 
Processing and Analysis Center/California Institute of Technology, funded by 
the National Aeronautics and Space Administration and the National Science 
Foundation. This work is partly supported by a Grant-in-Aid for JSPS Fellows 
for Young Researchers (MS) and for Scientific Research 23540265 (YU). 

{\it Facilities:} \facility{{\it Suzaku}}, \facility{{\it IRSF}}, 
\facility{{\it Kanata}}.

\end{document}